\documentclass[aps,prd,amsmath,amssymb,superscriptaddress,twocolumn,nofootinbib,showpacs,preprintnumbers,notitlepage]{revtex4-1}

\usepackage{color,paralist}
\usepackage{mathrsfs}
\usepackage{amsfonts,amsmath,amssymb,bm}
\usepackage{epsfig}
\usepackage{graphicx}
\usepackage{amsfonts}
\usepackage{slashed}
\usepackage{float}
\usepackage{cancel}
\usepackage{relsize}
\usepackage[dvipsnames]{xcolor}
\usepackage[colorlinks, citecolor=blue, linktoc=all, linkcolor=blue, urlcolor=blue]{hyperref}

\newcommand{\nn}{\nonumber}
\newcommand{\xb}{x}
\newcommand{\RLT}{R_{\mbox{\scriptsize \it LT}}}

\newcommand{\beq}{\begin{equation}}
\newcommand{\eeq}{\end{equation}}

\newcommand{\ber}{\begin{eqnarray}} 
\newcommand{\eer}{\end{eqnarray}}

\newcommand{\jlab}{Thomas Jefferson National Accelerator Facility, Newport News, Virginia 23606, USA}
\newcommand{\iu}{Center for Exploration of Energy and Matter, Indiana University, Bloomington, Indiana 47403, USA}

\newcommand{\msu}{Skobeltsyn Nuclear Physics Institute and Physics Department at Moscow State University, 119899 Moscow, Russia}
\newcommand{\rome}{INFN Sezione di Roma, P.le Aldo Moro 5, I-00185 Roma, Italy}

\begin{document}

\preprint{JLAB-THY-21-3369}
\title{Resonant contributions to inclusive nucleon structure functions \\ from exclusive meson electroproduction data}
\author{A.~N.~\surname{Hiller Blin}}
\affiliation{\jlab}
\author{W.~\surname{Melnitchouk}}
\affiliation{\jlab}
\author{V.~I.~\surname{Mokeev}}
\affiliation{\jlab}
\author{V.~D.~\surname{Burkert}}
\affiliation{\jlab}
\author{V.~V.~\surname{Chesnokov}}
\affiliation{\msu}
\author{A.~\surname{Pilloni}}
\affiliation{\rome}
\author{A.~P.~\surname{Szczepaniak}}
\affiliation{\iu}
\affiliation{\jlab}

\begin{abstract}
Nucleon resonance contributions to the inclusive proton $F_2$ and $F_L$ structure functions are computed from resonance electroexcitation amplitudes in the mass range up to 1.75~GeV extracted from CLAS exclusive meson electroproduction data.
Taking into account for the first time quantum interference effects, the resonance contributions are compared with inclusive proton structure functions evaluated from $(e,e'X)$ cross section data and the longitudinal to transverse cross section ratio.
Contributions from isospin-1/2 and 3/2 resonances remain substantial over the entire range of photon virtualities $Q^2 \lesssim 4$~GeV$^2$, where their electroexcitation amplitudes have been obtained, and their $Q^2$ evolution displays pronounced differences in the first, second and third resonance regions.
We compare the structure functions in the resonance region with those computed from parton distributions fitted to deep-inelastic scattering data, and extrapolated to the resonance region, providing new quantitative assessments of quark-hadron duality in inclusive electron-proton scattering.
\end{abstract}

\date{\today}
\maketitle

%%%%%%%%%%%%%%%%%%%%%%%%%%%%%%%%%%%%%%%%%%%%%%%%%%%%%%%%%%%%%%%%%%%%%%%%%%%%%
\section{Introduction}

Inclusive electron scattering from protons has for many years provided fundamental information on the structure of the nucleon.
Data on inclusive $(e,e'X)$ cross sections and structure functions $F_1$, $F_2$, and/or $F_L$ have been key ingredients in global QCD analyses of parton distribution functions (PDFs)~\cite{Harland-Lang:2014zoa, Dulat:2015mca, Ball:2017nwa, Moffat:2021dji, Accardi:2016qay, Sato:2019yez, Alekhin:2017kpj}.
The PDFs characterize the momentum distributions of the quark and gluon constituents of the nucleon, and can be systematically extracted from the structure functions via QCD factorization theorems~\cite{Collins:1989gx}.
The leading twist approximation is found to be accurate  at invariant masses of the final state hadrons $W \gtrsim 2$~GeV and photon virtualities $Q^2 \gtrsim 1-2$~GeV$^2$.
Consequently, global QCD analyses~\cite{Accardi:2016qay, Alekhin:2017kpj, Sato:2019yez} have traditionally made cuts in $W$ and $Q^2$ (or in the Bjorken scaling variable $\xb = Q^2/2M\nu$, where $M$ is the nucleon mass and $\nu = (W^2-M^2+Q^2)/2M$ is the energy transfer in the target rest frame) to avoid the resonance region, sometimes with even more conservative cuts of $W \gtrsim 3-3.5$~GeV~\cite{Harland-Lang:2014zoa, Dulat:2015mca, Ball:2017nwa, Moffat:2021dji} (for recent reviews see Refs.~\cite{Jimenez-Delgado:2013sma, Gao:2017yyd, Ethier:2020way}).

From another perspective, the need to understand strong interaction physics across a broad range of energy and distance scales has provided motivation to extend knowledge of inclusive structure functions covering wider regions of $W$ and $Q^2$, including the resonance region of $W < 2$~GeV and $Q^2 > 1$~GeV$^2$.
Moving into the nonperturbative regime, an important question is how low in $W$ and $Q^2$ one can go while still retaining a partonic interpretation of the scattering process.

There is indeed a long history of PDF computations from nonperturbative models of QCD~\cite{Parisi:1976fz, Jaffe:1974nj, LeYaouanc:1974da, Hughes:1977sr, Bell:1977wy, Benesh:1987ie, Signal:1989yc, Schreiber:1991qx, Schreiber:1991tc, Steffens:1993pc, Celenza:1982uk, Meyer:1990fr, Mulders:1992za, Barone:1993iq, Benesh:1993tc, Melnitchouk:1993nk, Melnitchouk:1994en, Kulagin:1995ia, Traini:1997jz, Cloet:2005pp, Cloet:2007em, Roberts:2013mja, Bednar:2018mtf, Chen:2020ijn}.
More recent approaches to this problem have focused on the extraction of PDFs from the lattice QCD calculation of nucleon matrix elements of nonlocal operators~\cite{Ji:2013dva, Radyushkin:2017cyf, Ma:2017pxb, Lin:2017snn, Bringewatt:2020ixn}, which in principle provide a systematic, though not necessarily straightforward, means of connecting PDFs obtained from global QCD analyses with those from the fundamental QCD theory.

The goal of bridging the strongly-coupled, nonperturbative realm with the  perturbative domain remains a challenge that drives ongoing endeavors.
Efforts to expand the range of $W$ (or, equivalently,~$\xb$) covered in global analyses to as low as $W$ = 1.75~GeV have been made, notably by the CTEQ-Jefferson Lab (CJ) Collaboration~\cite{Owens:2012bv, Accardi:2016qay}, in an effort to provide stronger constraints on the behavior of PDFs at large values of~$x$.
There are interesting theoretical predictions for the PDFs in the limit $x \to 1$~\cite{Feynman:1972xm, Farrar:1975yb, Close:1988br, Melnitchouk:1995fc, Nocera:2014uea}, which have never been quantitatively tested.
However, extracting information from data in this region requires careful treatment of subleading effects such as target mass corrections~\cite{Accardi:2016qay, Alekhin:2017kpj, Moffat:2019qll, Brady:2011uy} and higher twists, as well as factorization breaking corrections. 
Furthermore, at $W < 2$~GeV, the inclusive structure functions exhibit peaks related to the contributions from the excited states of the nucleon.
% $N^*$ and $\Delta^*$.
The electroexcitation amplitudes of the resonance should therefore be incorporated into the description of structure functions.

It was observed long ago by Bloom and Gilman~\cite{Bloom:1970xb} in pioneering experiments at SLAC that a remarkable ``duality'' exists between the structure functions in the nucleon resonance region, $W \lesssim 2$~GeV, when averaged over resonances, and the scaling function extrapolated from the deep-inelastic scattering (DIS) region at high~$W$ to the low-$W$ (or high-$x$) domain populated by resonances.  
In addition to the Bloom-Gilman duality, in the low-$x$ region the structure functions also exhibit the Veneziano duality between $s$ and $t$-channel resonance/Reggeon exchanges.

The early data have since been supplemented by high-precision measurements of inclusive electron scattering cross sections in the resonance region at Jefferson Lab Hall~B~\cite{Osipenko:2003bu, Prok:2014ltt} and Hall~C~\cite{Malace:2009kw, Christy:2007ve, Tvaskis:2016uxm, Liang:2004tj}.
The modern experimental studies confirmed and further elaborated on the observations of duality in unpolarized proton structure functions~\cite{Niculescu:2000tj, Niculescu:2000tk, Malace:2009kw, Liuti:2001qk, Niculescu:2015wka}.
A compilation of the data for unpolarized structure functions and inclusive cross sections in the range $1.07 \leq W \leq 2$~GeV and $0.5 \leq Q^2 \leq 7$~GeV$^2$, together with a tool for the interpolation between bins, is available online~\cite{Golubenko:2019gxz, CLAS:DB, CLAS:SFDB}.
Subsequently, studies of Bloom-Gilman--type duality have been extended to various other observables, including unpolarized neutron structure functions~\cite{Malace:2009dg}, spin-dependent nucleon structure functions~\cite{Bosted:2006gp}, $\gamma^* p$ helicity cross sections~\cite{Malace:2011ad}, neutrino scattering~\cite{Lalakulich:2006yn, Lalakulich:2008tu}, and nuclear structure functions~\cite{Arrington:2003nt, Melnitchouk:2001nr}.
Models exist that account for both the Veneziano and Bloom-Gilman dualities~\cite{Jenkovszky:2012dc}, but a quantitative understanding is still missing.

It has long been realized that, when integrating the structure functions over $W$ (or $\xb$), duality can be related to QCD through the operator product expansion (OPE).
Here, the moments of the structure functions are expanded in inverse powers of the hard scale $Q^2$, with numerators given by matrix elements of local quark-gluon operators characterized by a certain twist (mass dimension minus spin).
The leading term, referred to as leading twist, involves quark and gluon bilinear operators, and is associated with incoherent scattering with individual partons in the nucleon.
The correction terms, referred to as higher twist, are determined by matrix elements of multi-parton operators, and capture elements of long-distance, nonperturbative quark-gluon dynamics associated with color confinement in QCD~\cite{Ji:1994br}.
In this language the appearance of duality is interpreted in terms of the dominance of the leading twist term and suppression of higher twist contributions to the structure function moments~\cite{DeRujula:1976baf}.

Moreover, it was later shown that this interpretation could be generalized to truncated moments of structure functions, involving finite intervals of $x$, without the need to extrapolate outside of measured kinematics~\cite{Forte:1998nw, Forte:2000wh}.
For the leading twist part of the structure functions, the truncated moments were found to obey DGLAP-type $Q^2$ evolution equations~\cite{Piccione:2001vf, Kotlorz:2006dj, Kotlorz:2016icu}, so that deviations of the empirical truncated moments from the predicted $Q^2$ behavior could reveal the magnitude of the higher twist contributions.
An earlier phenomenological analysis~\cite{Psaker:2008ju} of Jefferson Lab Hall~C data~\cite{Liang:2004tj} found deviations of the truncated moments of the resonance region data ($W \leq 2$~GeV) from leading twist behavior 
of $\lesssim 15\%$ for $Q^2 > 1$~GeV$^2$.
For the individual resonance regions, the first ($\Delta$) resonance region was found with $\approx -15\%$ higher twist contribution at $Q^2 \sim 1$~GeV$^2$, while the second and third resonance regions had somewhat larger duality violations, $\approx -15\%$ to $\approx +25\%$ and $\approx 0$ to $\approx +15\%$ in the range $Q^2 \sim 1-5$~GeV$^2$, respectively.

While the integrated version of the Bloom-Gilman duality can formally be framed within QCD, understanding the functional dependence on $x$ is more challenging and requires  nonperturbative model arguments.
Indeed, the question of how to obtain a smooth, scaling function from a sum of sharp resonances has inspired considerable theoretical attention~\cite{Domokos:1971ds, Domokos:1971bw, Domokos:1972yc, Close:2001ha, Isgur:2001bt, Jeschonnek:2002db, Close:2003wz, Close:2009yj} (see Ref.~\cite{Melnitchouk:2005zr} for additional references to the literature).
Phenomenological calculations exploring how exclusive contributions to the final states build up the inclusive structure functions have been made by modelling the resonant contributions~\cite{Davidovsky:2002nj}, as well as combining resonance models with Regge physics~\cite{Fiore:2002re, Jenkovszky:2012dc}.
Attempts to correlate the behavior of individual resonances with $Q^2$ with the $x$ dependence of the dual leading twist structure function have been made~\cite{DeRujula:1976baf}, including for the super-local case of elastic scattering at $x \sim 1$, with interesting phenomenological predictions~\cite{Simula:1999rr, Ent:2000jj, Melnitchouk:2001eh}.

In the past decade, the experimental program exploring exclusive $\pi^+ n$, $\pi^0 p$, $\eta p$, and $\pi^+ \pi^-p$ electroproduction channels in the resonance region with CLAS at Jefferson Lab has provided important new information on the nucleon resonance electroexcitation amplitudes, or $\gamma^* p N^*$ electrocouplings.
These include the electrocouplings of most nucleon resonances in the mass range $W \leq 1.75$~GeV and $Q^2 \leq 5$~GeV$^2$~\cite{Aznauryan:2011qj, Aznauryan:2009mx, Mokeev:2012vsa, Mokeev:2015lda, Park:2014yea, Carman:2020qmb}, as well as the new baryon state $N'(1720)\,3/2^+$ observed in combined studies of the $\pi^+ \pi^- p$ photo- and electroproduction data~\cite{Mokeev:2020hhu}.
The consistency of the results for the $\gamma^* p N^*$ electrocouplings of several nucleon resonances from independent studies of $\pi N$ and $\pi^+ \pi^- p$ electroproduction allows us to determine the uncertainties related to the use of the reaction models in the extraction of these quantities~\cite{Mokeev:2012vsa, Mokeev:2015lda}. 
These results from CLAS make it possible to evaluate the resonant contributions to inclusive electron scattering in the resonance region using parameters of the individual nucleon resonances extracted from data, pioneered in Ref.~\cite{Blin:2019fre}.

The majority of previous studies of inclusive $(e,e'X)$ processes have employed parameterizations of the ratio $\RLT$ of the longitudinal to transverse virtual photon cross sections from DIS region data.
However, dedicated studies in Jefferson Lab Hall~C have provided information on $\RLT$ from data in the resonance region, based on Rosenbluth separation~\cite{Tvaskis:2016uxm, Liang:2004tj}. 
Using the Hall~C results for $\RLT$, we firstly update the determination of the inclusive $F_2$ and $F_L$ structure functions from cross section data.
These are compared with the resonant contributions to the structure functions computed from the empirical $\gamma^* p N^*$ electrocouplings obtained from the analyses of exclusive meson electroproduction data from CLAS~\cite{Carman:2020qmb,Aznauryan:2011qj}.
We update the {\it ansatz} of Ref.~\cite{Blin:2019fre} for the evaluation of the resonant contributions, allowing interference effects to be included between the excited nucleon states, with the overall resonant amplitudes expressed as a coherent sum over all relevant resonances in the mass range $W < 1.75$~GeV.
In particular, our analysis enables us to assess the role of the interference between resonances in the composition of the inclusive structure functions.

The availability of phenomenological results on $\gamma^*pN^*$ electrocouplings allows us to quantitatively explore the evolution of the resonant contributions to the inclusive electron scattering observables with $Q^2$.
The experimental data on the $F_2$ and $F_L$ structure functions elucidate the role of the different nucleon excited states in the generation of the inclusive electron scattering cross sections, and their contributions to the truncated moments of the structure functions in the first, second, and third resonance regions.

Ultimately, the information on the $Q^2$ evolution of the resonant contributions over a broad range of $W < 1.75$~GeV can shed light onto the prospects for the studies of the $\gamma^* p N^*$ electrocouplings at $Q^2 > 5$~GeV$^2$ with the CLAS12 detector at Jefferson Lab~\cite{Carman:2020qmb, Brodsky:2020vco}. 
Furthermore, the data on the inclusive structure functions, combined with the resonance contributions presented here, may provide insights into nucleon PDFs at large $\xb$ values in the DIS-resonance transition region.

The rest of the paper is organized as follows. 
In Sec.~\ref{sec:RLT} we give a brief review of the main definitions of structure functions and cross sections in inclusive electron-proton scattering, where we also summarize the status of $\RLT$ measurements.
The formulas for the resonant contributions to structure functions are presented in Sec.~\ref{sec:res_eval}, in terms of the $\gamma^*pN^*$ electrocouplings for transverse and longitudinal photons.
Results on the resonant contributions to inclusive $F_2$ and $F_L$ structure functions are discussed in Sec.~\ref{sec:results}, where we compare individual and overall resonance contributions to the structure function data.
In Sec.~\ref{sec:duality} we compare the structure functions in the resonance region with those extrapolated from the high-$W$ region, to critically examine the degree to which quark-hadron duality is satisfied by the data, for both the $x$-dependent functions and their lowest truncated moments. 
Finally, in Sec.~\ref{sec:outlook} we summarize our results and discuss future extensions and applications of our work.
In Appendix~\ref{app:TMC}, for completeness we collect the main formulas for target mass corrections from the OPE and collinear factorization approaches.

%%%%%%%%%%%%%%%%%%%%%%%%%%%%%%%%%%%%%%%%%%%%%%%%%%%%%%%%%%%%%%%%%%%%%%%%%%%%%
\section{Structure functions from inclusive electron scattering}
\label{sec:RLT}

We begin the discussion here by firstly reviewing the basic formulas for cross sections and structure functions for inclusive electron-nucleon scattering.
Following this, we re-evaluate the $F_2$ and $F_L$ structure functions from inclusive proton cross section data from CLAS~\cite{Osipenko:2003bu}, using the most recent information on the ratio of longitudinal to transverse cross sections from Jefferson Lab Hall~C measurements~\cite{Liang:2004tj, Christy:2007ve, Tvaskis:2016uxm}, which will be used in our subsequent analysis.

The inclusive $F_1$ and $F_2$ structure functions are related to the total virtual photon--nucleon scattering cross sections $\sigma_T$ and $\sigma_L$, for transversely and longitudinally polarized photons, respectively~\cite{Drechsel:2002ar},
\begin{subequations}
\label{sf} 
\begin{align}
F_1(W,Q^2) &=\frac{{K M}}{4\pi^2\alpha}\, \sigma_T(W,Q^2),  \\
F_2(W,Q^2) &=\frac{{K M}}{4\pi^2\alpha}\, \frac{2\xb}{\rho^2}
             \left(\sigma_T(W,Q^2) +\sigma_L(W,Q^2)\right),
\end{align}
\end{subequations}
where $\alpha$ is Sommerfeld's fine structure constant, $K$ is the equivalent photon energy, defined here in the Hand convention as
\begin{align}
K&=\frac{W^2-M^2}{2M},
\end{align}
and $\rho$ is a kinematic parameter,
\begin{align}
\rho^2 = 1 + \frac{4 M^2 \xb^2}{Q^2}.
\label{rhopar}
\end{align}
The $F_2$ structure function can also be written in terms of the unpolarized virtual photoproduction cross section~$\sigma_{U}$,
\begin{align}
\label{eq:F2sig}
F_2(W,Q^2)
& = \frac{K M}{4\pi^2\alpha} 
    \frac{2\xb}{\rho^2} 
    \frac{1+\RLT}{1+\epsilon \RLT}\, \sigma_{U}(W,Q^2),
\end{align}
where
\begin{equation}
\label{xs}
\sigma_{U}(W,Q^2) = \sigma_T(W,Q^2) + \epsilon\, \sigma_L(W,Q^2),
\end{equation}
and $\epsilon$ is the degree of transverse virtual photon polarization, determined by the scattered electron angle, $\theta_e$, in the laboratory frame and the parameter $\rho$ in Eq.~(\ref{rhopar}),
\begin{equation}
\epsilon = \left( 1 + \frac{2 \rho^2}{\rho^2-1} \tan^2\frac{\theta_e}{2}
           \right)^{-1}.
\label{SigmaU1}
\end{equation}
In Eq.~(\ref{eq:F2sig}) $\RLT$ is the ratio of longitudinal to transverse virtual photon cross sections,
\begin{equation}
\label{rlt}
\RLT(W,Q^2)=\frac{\sigma_{L}(W,Q^2)}{\sigma_{T}(W,Q^2)}.
\end{equation}
{It is also convenient to} define the longitudinal structure function, $F_L$, in terms of the longitudinal cross section $\sigma_L$, or, equivalently, in terms of the $F_1$ and $F_2$ structure functions,
\begin{align}
F_L(W,Q^2) &=\frac{K M}{4\pi^2\alpha} 
    2\xb\,\sigma_L(W,Q^2) \nn \\
&= \rho^2 F_2(W,Q^2) - 2 x F_1(W,Q^2).
\label{sfl}
\end{align}

\begin{figure}[t]%[H]
\begin{center}
\includegraphics[width=.45\textwidth]{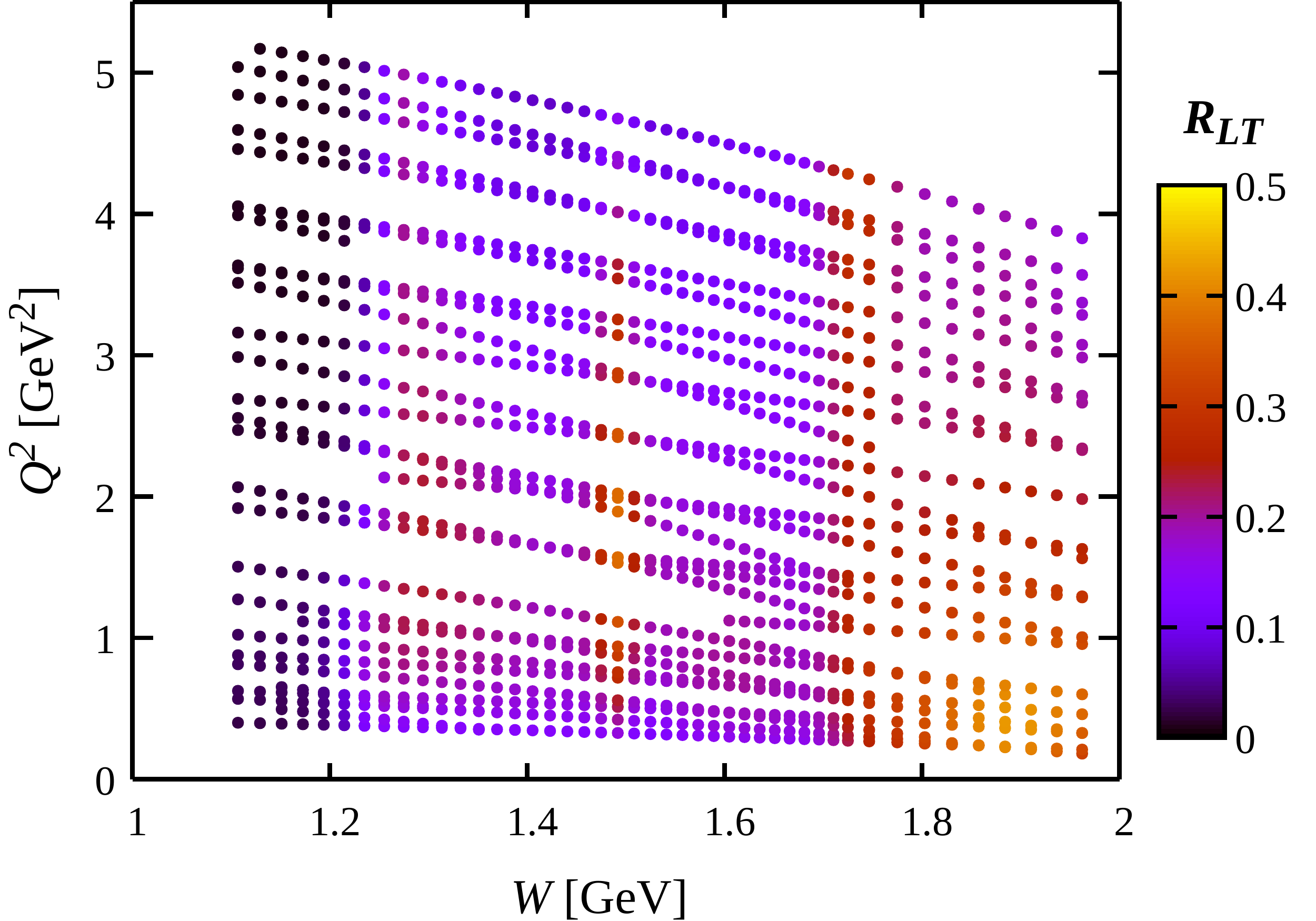}
\end{center}
\vspace*{-0.5cm}
\caption{Kinematic coverage in $W$ and $Q^2$ of $\RLT$ measurements from Jefferson Lab Hall~C data, as presented in Ref.~\cite{Liang:2004tj}.}
\label{F:RLT1}
\end{figure}

While the total inclusive cross sections have been measured in many experiments across a large range of kinematics, the extraction of the individual $\sigma_T$ and $\sigma_L$ components, using the Rosenbluth separation technique, is considerably more difficult, and the $\RLT$ ratio is known to much less accuracy.
The experiments from Hall~C at Jefferson Lab~\cite{Liang:2004tj, Christy:2007ve, Tvaskis:2016uxm} provided the first data on $\RLT$ in the resonance region.
The kinematic coverage of the data from the Jefferson Lab E94-110 experiment is shown in Fig.~\ref{F:RLT1}, where we also indicate the interpolated values of $\RLT$ from the analysis of Ref.~\cite{Liang:2004tj}.
Using the recent results on $\RLT$~\cite{Liang:2004tj, Christy:2007ve, Tvaskis:2016uxm}, we have extracted the $F_1$ and $F_2$ structure functions from the CLAS inclusive electron-proton scattering cross sections~\cite{Osipenko:2003bu} at $W<2$~GeV and at $Q^2<3.5$~GeV$^2$, updating the results from Ref.~\cite{Blin:2019fre}.
In addition, we perform a further interpolation and extrapolation in order to obtain a finer grid over $W$ and $Q^2$, covering the range of $Q^2<7$~GeV$^2$.
However, the extrapolation region is not used in the present analysis, which we restrict to the actual experimental region, where the results on $\gamma^*pN^*$ electrocouplings are available~\cite{Blin:2019fre, Carman:2020qmb}.

\begin{figure}[t]%[H]
\begin{center}
\includegraphics[width=.45\textwidth]{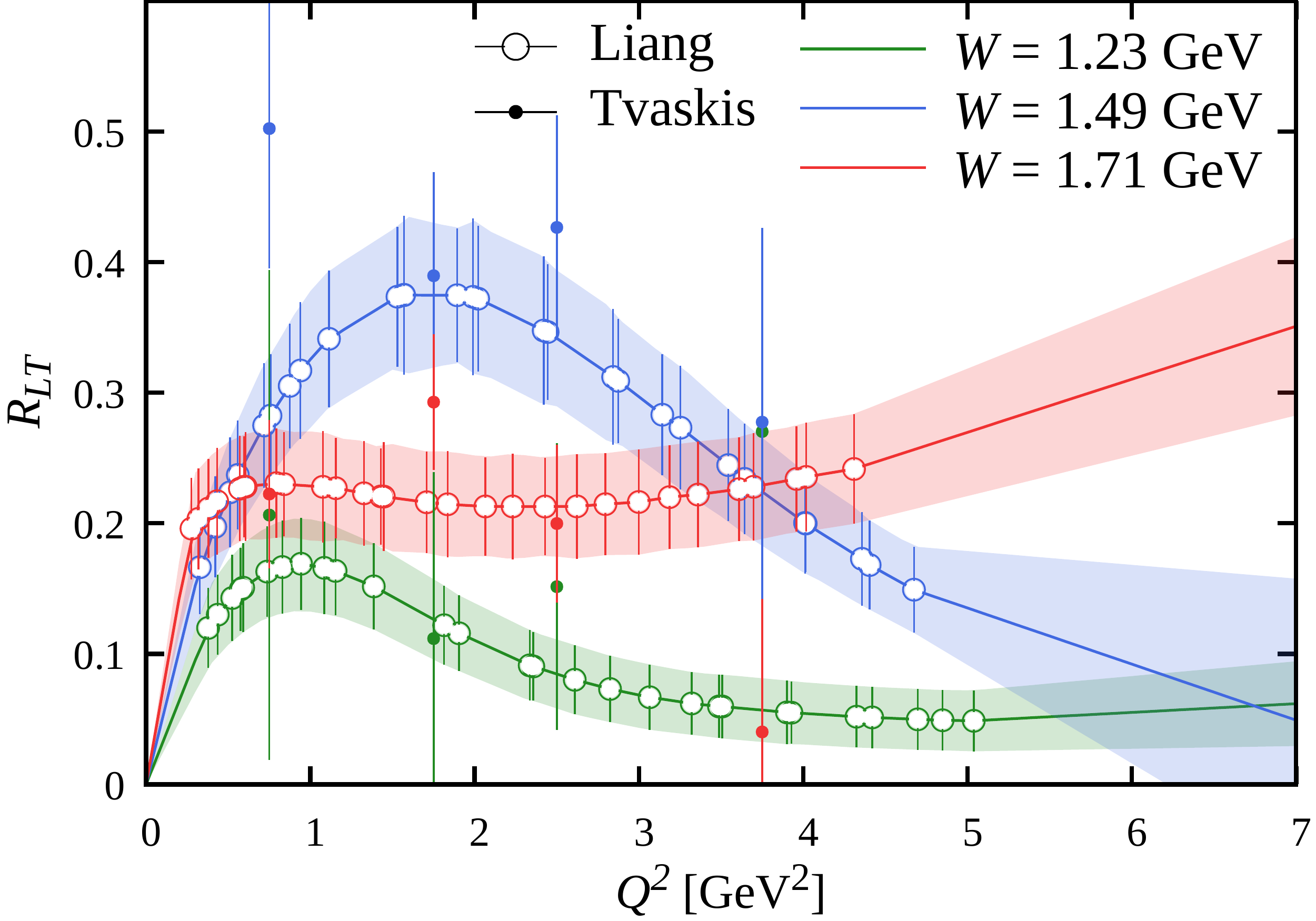}
\end{center}
\vspace*{-0.5cm}
\caption{Longitudinal to transverse cross section ratio $\RLT$ versus $Q^2$ in several bins of $W$, represented as interpolation/extrapolation bands computed from the data in Ref.~\cite{Liang:2004tj} (large open circles) and compared with the Hall~C data from Ref.~\cite{Tvaskis:2016uxm} (small filled circles).}
\label{F:RLT2}
\end{figure}

In Fig.~\ref{F:RLT2} representative examples of $\RLT$ are shown as a function of $Q^2$ in several fixed bins of $W$.
Extrapolation to larger $Q^2$ values is performed by fitting a second degree polynomial to the grid of Ref.~\cite{Liang:2004tj} at $Q^2 > 3$~GeV$^2$. 
For the error estimate at $Q^2 = 7$~GeV$^2$, we choose the largest of the following two approaches: 
    (i) taking the difference between fits obtained using second and first degree polynomials;
    (ii) extrapolating the error of the last $Q^2$ bin in Ref.~\cite{Liang:2004tj} to $Q^2 = 7$~GeV$^2$ by scaling it with the extrapolated value using the second degree polynomial at $Q^2 = 7$~GeV$^2$.
The full uncertainty band is then obtained as the interpolation between the data errors and the larger of the two estimated errors at $Q^2=7$~GeV$^2$. 
At $Q^2=0$ we set the ratio $\RLT$ to zero.
The comparison in Fig.~\ref{F:RLT2} illustrates the differences between the $Q^2$ evolution of our $\RLT$ interpolation bands and those of the grids in the analyses of Refs.~\cite{Liang:2004tj, Tvaskis:2016uxm} evaluated at the same values of $W$.

We stress that because of the almost {complete,} $4\pi$ angular acceptance of the CLAS detector, the data from Ref.~\cite{Osipenko:2003bu} span the entire kinematically allowed range of $W$ in each bin of $Q^2$, with a bin width $\Delta Q^2=0.05$~GeV$^2$.
The broad coverage over $W$ is particularly important in the resonance region of $W < 2$~GeV due to the presence of resonance structures in the experimental data at these kinematics.
Outside of the kinematic area covered by the data of Ref.~\cite{Osipenko:2003bu}, at $Q^2 > 3.5$~GeV$^2$, we use the parameterization of the world data developed in Refs.~\cite{Golubenko:2019gxz, Christy:2007ve}.
The data used in those analyses had more limited coverage over $W$ at any given $Q^2$ than those in~\cite{Osipenko:2003bu}.
However, at the higher values $Q^2 > 3.5$~GeV$^2$ the resonance structures become less pronounced, which makes the data interpolation over $W$ and $Q^2$ more reliable.

%%%%%%%%%%%%%%%%%%%%%%%%%%%%%%%%%%%%%%%%%%%%%%%%%%%%%%%%%%%%%%%%%%%%%%%%%%%%%
\section{Resonant contributions to unpolarized inclusive structure functions}
\label{sec:res_eval}

In this section, we describe the evaluation of the resonant contributions to the inclusive structure functions $F_1$ and $F_2$, or their combination, $F_L$.
The formalism used in the analysis of unpolarized structure functions in Ref.~\cite{Blin:2019fre} has been updated in the present paper, allowing us to explicitly take into account interference effects between different resonances.

The contributions of the isospin-1/2 $N^*$ and isospin-3/2 $\Delta^*$ resonances to inclusive $\gamma^*p$ cross sections are described~\cite{Blin:2019fre} by employing $\gamma^* p N^*$ electrocouplings from exclusive meson
% the CLAS $\pi N$, $\eta p$, and $\pi^+ \pi^- p$ 
electroproduction data~\cite{Aznauryan:2009mx, Aznauryan:2011qj, Mokeev:2012vsa, Mokeev:2016hqv, CLAS:coupsDB, Park:2014yea, CLAS:coups, Carman:2020qmb}.
Currently, this information is limited to the resonances in the mass range up to $W \approx 1.75$~GeV.
Apart from the well-established, four-star resonances identified in the Review of Particle Physics (RPP)~\cite{Zyla:2020zbs}, we include also the recently observed $N'(1720)~3/2^+$ state~\cite{Mokeev:2020hhu}, since, as discussed below, this state plays an important role in the generation of the peak in the third resonance region of the inclusive electron scattering data.
A list of the nucleon resonances included in our analysis, together with their properties, can be found in Ref.~\cite{Blin:2019fre}.

If the interference between different resonances is neglected, the contribution from a resonance $R$ of mass $M_R$ and spin $J_R$ to the transverse ($\sigma_{T}^R$) and longitudinal ($\sigma_{L}^R$) inclusive virtual photon-proton cross sections is given by an incoherent sum over Breit-Wigner cross sections from all  resonances~\cite{Mokeev:2012vsa,Blin:2019fre},
\begin{align}
\label{Eq:BW}
&\sigma_{T,L}^R(W,Q^2)& \nn\\
&=\frac{\pi}{q_\gamma^2} \sum_{R}(2J_R+1) 
  \frac{M_R^2\, \Gamma_R (W)\, \Gamma^{T,L}_{\gamma,R}(M_R,Q^2)}
       {\big(M_R^2-W^2\big)^2 + \big(M_R\Gamma_R(W)\big)^2}, 
\end{align}
where $\Gamma_R (W)$ is the $W$-dependent  total decay width of the resonance $R$ ~\cite{Blin:2019fre}. 
Here, $q_\gamma$ and $E_\gamma$ are the virtual photon three-momentum and energy in the photon-target proton center of mass frame, respectively,
\begin{subequations}
\begin{align}
q_\gamma &= \sqrt{Q^2+E^2_\gamma},     &
E_\gamma &= \frac{W^2-Q^2-M^2}{2W}.
\end{align}
\end{subequations}
The electromagnetic decay widths $\Gamma^{T,L}_{\gamma,R}$ of the resonance $R$ to the $\gamma^* p$ state with transversely or longitudinally polarized virtual photons are related to the electrocouplings $A_{1/2}^R(Q^2)$, $A_{3/2}^R(Q^2)$ and $S_{1/2}^R(Q^2)$ by 
\begin{subequations}
\begin{align}
\Gamma_{\gamma, R}^T(W=M_R,Q^2)
&= \frac{q^2_{\gamma,R}(Q^2)}{\pi} \frac{2M}{(2J_R+1)M_R}   \nn\\
& \hspace*{-1cm} \times 
   \Big( \big|A_{1/2}^R(Q^2)\big|^2 + \big|A_{3/2}^R(Q^2)\big|^2
   \Big),       
\\
\Gamma_{\gamma, R}^L(W=M_R,Q^2)
&= \frac{2 q^2_{\gamma,R}(Q^2)}{\pi} \frac{2M}{(2J_R+1)M_R}  \nn\\
& \hspace*{-1cm} \times \big|S_{1/2}^R(Q^2)\big|^2,
\label{Eq:EMWidths}
\end{align}
\end{subequations}
where $q_{\gamma,R} \equiv q_{\gamma}(W\!=\!M_R)$ is the virtual photon momentum at the resonance peak. 
Further details about the resonance electromagnetic decay widths $\Gamma_{\gamma, R}^{T(L)}$ and total decay widths $\Gamma_R(W)$, can be found in Ref.~\cite{Blin:2019fre}.
The resonant contributions to the inclusive structure functions are then obtained by inserting the resonant cross sections in Eq.~(\ref{Eq:BW}) into Eqs.~(\ref{sf}).

In order to take into account the interference between different resonances, the  resonant contribution to the structure functions needs to be evaluated in terms of a coherent sum of $\gamma^* p \to N^*, \Delta^* \to X$ amplitudes, where $X$ stands for all final states populated in the resonance decays.
The contribution from each resonance of spin $J$, isospin $I$, and parity $\eta$ can be described by the amplitudes $G_{m}^R$, where $m = +1, 0, -1$ is the virtual photon spin projection onto the quantization axis $Oz$, aligned along the direction of the virtual photon momentum.
Adding the amplitudes coherently, the resonant contributions to the structure functions can then be written as~\cite{Carlson:1998gf, Melnitchouk:2005zr}
\begin{subequations}
\label{Eq:coherent}
\begin{eqnarray}
F_1^R
&=& M^2 \sum_{IJ\eta} 
    \Bigg[
    \bigg| \sum_{R^{IJ\eta}} G_+^{R^{IJ\eta}} \bigg|^2
  + \bigg| \sum_{R^{IJ\eta}} G_-^{R^{IJ\eta}} \bigg|^2
    \Bigg], \qquad
\\
\rho^2 F_2^R 
&=& M \nu \sum_{IJ\eta}
    \Bigg[
    \bigg| \sum_{R^{IJ\eta}} G_+^{R^{IJ\eta}} \bigg|^2 +  \bigg| \sum_{R^{IJ\eta}} G_-^{R^{IJ\eta}} \bigg|^2
\nonumber\\
& & \qquad
 +\ 2\, \bigg| \sum_{R^{IJ\eta}} G_0^{R^{IJ\eta}} \bigg|^2
    \Bigg],
\\
F_L^R 
&=& \rho^2 F_2^R - 2\xb F_1^R,
\end{eqnarray}
\end{subequations}
where the outer sum runs over the possible values of spin $J$, isospin $I$ and intrinsic parity $\eta$, and the inner sums run over all those resonances $R^{IJ\eta}$ which satisfy $J_R=J$, $I_R=I$ and $\eta_R=\eta$ for the spin, isospin and parity of the resonance $R$.
As detailed below, the combination of inner and outer sums in Eqs.~(\ref{Eq:coherent}) reflects the cancellation between interference terms for resonances of different spin, isospin or parity after integration over the final hadron emission angles in the center of mass frame, and after the sum over the resonance decays into all possible final states.
Ultimately, this amounts to taking into account only the interference between states of the same isospin, spin and parity, which in practice involves the pairs of excited nucleon states $N(1440)\,1/2^+$ and $N(1710)\,1/2^+$,  $N(1535)\,1/2^-$ and $N(1650)\,1/2^-$, and $N(1720)\,3/2^+$ and $N'(1720)\,3/2^+$.

The amplitudes $G_{m}^R$~\cite{Carlson:1998gf, Melnitchouk:2005zr} in Eqs.~(\ref{Eq:coherent}) are proportional to the electrocouplings $A_{1/2}^R$, $A_{3/2}^R$, and $S_{1/2}^R$~\cite{Aznauryan:2011qj}.
The $G_{+}^R$ and $A_{1/2}^R$ amplitudes are defined for the same values of the virtual photon and target proton spin projections onto $Oz$, while the $G_{0}^R$ and $G_{-}^R$ amplitudes are obtained from $S_{1/2}^R$ and $A_{3/2}^R$ after space reflection,
\begin{subequations}
\begin{eqnarray}
G_{+}^R &\sim& A_{1/2}^R,           \\
G_{0}^R &\sim& S_{1/2}^R\, (-1)^{P},   \\
G_{-}^R &\sim& A_{3/2}^R\, (-1)^{P},
\end{eqnarray} \label{eq:ga}
\end{subequations}
where the parity transformation factor $P$ is given by
\begin{equation}
\label{eq:transform}
P = \frac{\eta}{\eta_\gamma\eta_N}(-1)^{J-J_\gamma-J_N}
  = \eta(-1)^{J-1/2}.
\end{equation} 
The subscripts $\gamma$ and $N$ here denote the virtual photon and nucleon, respectively.
Note that for unpolarized structure functions, as considered in the present analysis, the phase factor in Eq.~(\ref{eq:transform}) is not relevant for the absolute values of the $G_+^{R^{IJ\eta}}$ amplitudes in Eqs.~(\ref{Eq:coherent}).
The approach described above is also applicable for the evaluation of the resonant contributions to polarized inclusive structure functions, where this phase factor does become relevant.
The contribution from a single resonance $R$ of finite and $W$-dependent decay width $\Gamma_R(W)$ is then evaluated by expressing the amplitudes $G_{m}^R$ computed within the Breit-Wigner {\it ansatz} of Ref.~\cite{Mokeev:2012vsa},
\begin{subequations}
\label{gm_amplitudes}
\begin{align}
G_+^R &= C \frac{\sqrt{M_R\Gamma_R(W)}}{M_R^2-W^2-i\Gamma_R(W)M_R}\, A^R_{1/2}(Q^2),
\\
G_-^R &= C \frac{\sqrt{M_R\Gamma_R(W)}}{M_R^2-W^2-i\Gamma_R(W)M_R}\, A^R_{3/2}(Q^2)(-1)^P,
\\
G_0^R &= C \frac{\sqrt{M_R\Gamma_R(W)}}{M_R^2-W^2-i\Gamma_R(W)M_R}\, S^R_{1/2}(Q^2)(-1)^P,
\end{align}
\end{subequations}
where $C$ denotes the conversion factor that transforms the amplitudes $G_{m}^R$ into the convention corresponding to the Breit-Wigner cross section of Eq.~(\ref{Eq:BW}) for a single resonance contribution.
{
The factor $C$ can be evaluated by computing the structure functions in Eqs.~(\ref{sf}) using the cross sections in Eq.~(\ref{Eq:BW}), and comparing with the structure functions computed from Eqs.~(\ref{Eq:coherent}) with the $G^{R}_{m}$ amplitudes defined in Eqs.~(\ref{gm_amplitudes}) for the contribution from a single resonance $R$.
Setting the resulting expressions to be equal, the conversion factor is determined to be}
\begin{align}
\label{conversion_factor}
   C &= \frac{1}{4\pi} \sqrt{\frac{W^2-M^2}{\alpha M}} \frac{q_{\gamma, R}}{q_\gamma}.
\end{align}
The resulting resonance interference can be constructive or destructive, since the electrocouplings can be positive-valued or negative-valued real numbers.

It is well known that resonances with different quantum numbers do not interfere in the $W$ dependence of the integrated cross sections.
{
This can be seen explicitly % as detailed in the following.
by observing that the resonant contributions to $F_2$ and $F_L$ described by Eqs.~(\ref{Eq:coherent}) with the amplitudes of Eqs.~(\ref{gm_amplitudes})} are obtained after integration over the center of mass emission angle of one of the final hadrons in all exclusive channels.
The angular dependence of the amplitudes in Eqs.~\eqref{gm_amplitudes} is implicitly given by the Wigner functions $d^{J}_{\nu\mu}(\theta)$, where $\theta$ is the polar angle between the virtual photon and the meson directions in the center of mass frame, and $\nu$, $\mu$ are the sum of spin projections of the initial virtual photon and proton onto the virtual photon direction, and sum of spin projections of the final state particles onto the direction of one of the final particles, respectively.
The structure functions defined in Eqs.~\eqref{Eq:coherent} contain the amplitude squared integrated over $\theta$. 
When the integral is performed, the interference terms between resonances of different spins $J$ vanish because of the orthogonality of the Wigner functions. 
Similarly, the interference terms between resonances of different parities vanish due to the orthogonality of the eigenvectors of the orbital angular momentum operator, upon integration over $\theta$. 
Finally, the interference terms between resonances of different isospin vanish after summing over the decays to the final states with  all possible isospin projections, because of the orthogonality relations for the isospin Clebsch-Gordan coefficients.
Therefore, only interference terms from the resonances of the same spin $J$, isospin $I$ and parity $\eta$ contribute to the observables of inclusive processes. 

One needs to take particular care with interferences between resonances of close mass, such as the  $N(1720)\,3/2^+$ and the $N'(1720)\,3/2^+$. 
The recent CLAS analysis~\cite{Mokeev:2020hhu} found  that the decays into the $\pi \Delta$ and $\rho\,p$ final states account for the largest part of the hadronic width for these resonances, and the interference between these two resonances should be evaluated in each of these exclusive channels separately. 
In our {\it ansatz} of Eq.~\eqref{gm_amplitudes}, we do not take into account differences in hadronic decay widths of the $N'(1720)\,3/2^+$ and $N(1720)\,3/2^+$ resonances into the $\pi\Delta$ and $\rho p$ final states. 
While this is approximately the case for the decay into the $\pi\Delta$ final state, where the branching fraction for the $N'(1720)\,3/2^+$ is only about 20\% larger than that of the $N(1720)\,3/2^+$, the differences in the $\rho\,p$ channel are too large to be neglected: the branching fraction for the $N(1720)\,3/2^+$ is about 5.5 times larger than that of the $N'(1720)\,3/2^+$.
Therefore, compared to our {\it ansatz}, there is a suppression factor coming from differences in the two resonance decays widths both to the $\pi\Delta$ and the $\rho p$ final states. 
These suppression factors for the interference between the $N'(1720)\,3/2^+$ and $N(1720)\,3/2^+$ resonances in the $\pi \Delta$ and $\rho p$ channels are equal to the square root of the decay width ratio to the $\pi \Delta$ or $\rho p$ final state for the resonance with larger decay width over the resonance of smaller decay width. 
In order to estimate the effect on our {\it ansatz}, we average between the suppression factors for the $\pi \Delta$ and $\rho p$ final states:
\begin{align}
   & \frac12\left(
    \sqrt{\frac{\Gamma_{N'(1720)\to\pi\Delta}}{\Gamma_{N(1720)\to\pi\Delta}}}+
    \sqrt{\frac{\Gamma_{N(1720)\to\rho\, p}}{\Gamma_{N'(1720)\to\rho\, p}}}
    \right)
  \approx 1.72,
\end{align}
and use this value for the suppression of the interference term between the $N(1720)\,3/2^+$ and $N'(1720)\,3/2^+$ states.

%%%%%%%%%%%%%%%%%%%%%%%%%%%%%%%%%%%%%%%%%%%%%%%%%%%%%%%%%%%%%%%%%%%%%%%%%%%%%
\section{Numerical results}
\label{sec:results}

In this section we present the results of our calculations of the resonance contributions to the inclusive proton $F_2$ and $F_L$ structure functions from the experimental results on the $\gamma^*pN^*$ electrocouplings \cite{Blin:2019fre,Carman:2020qmb}, and compare these with the structure functions extracted from experimental inclusive cross sections from CLAS~\cite{Osipenko:2003bu} combined with empirical results on the $\RLT$ ratio from Hall~C~\cite{Liang:2004tj,Tvaskis:2016uxm}. The data are then interpolated on a grid of $(W,Q^2)$ values presented in our paper, by employing the web tool in Ref.~\cite{CLAS:SFDB}.

We examine the role of individual resonance contributions to the structure functions, and assess the importance of the interference effects computed from our coherent {\it ansatz} relative to the incoherent approach from Ref.~\cite{Blin:2019fre}.

%............................................................................
\subsection{Resonance contributions to structure functions}

\begin{figure*}[t]
\includegraphics[width=0.9\textwidth]{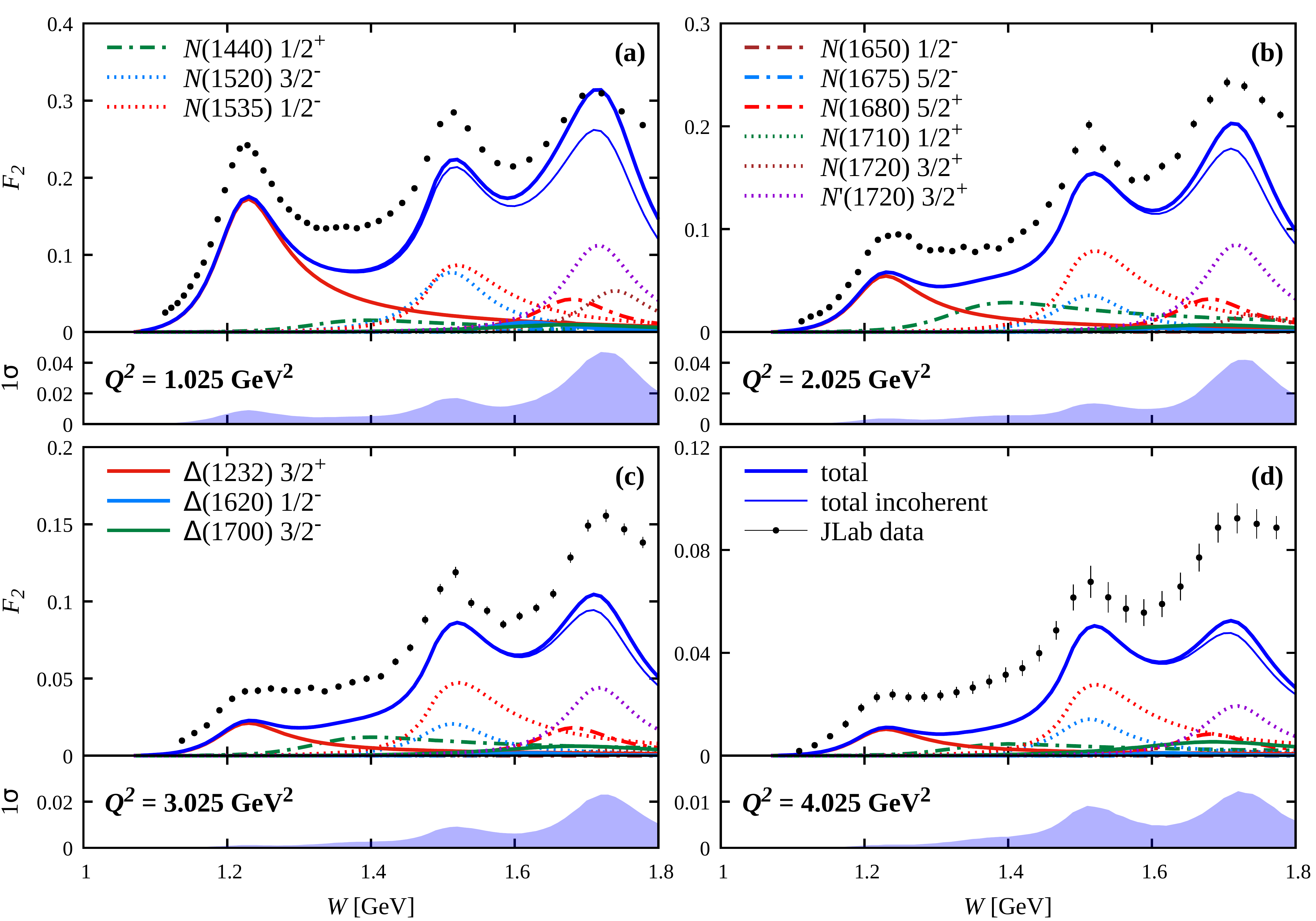}
\caption{Proton $F_2$ structure function at fixed values of $Q^2$ corresponding to the data bins in Ref.~\cite{Osipenko:2003bu}: 
{\bf (a)}~$Q^2=1.025$~GeV$^2$, 
{\bf (b)}~$Q^2=2.025$~GeV$^2$, 
{\bf (c)}~$Q^2=3.025$~GeV$^2$, 
{\bf (d)}~$Q^2=4.025$~GeV$^2$. The interpolated experimental data from Ref.~\cite{CLAS:SFDB} (filled black circles) are compared with the full resonant  contributions to the  structure functions computed by adding amplitudes (thick blue curves) and cross sections (thin blue curves) from the contributing resonances, using the central values of their electrocouplings. The contributions from individual resonances are shown separately, as indicated in the legends. Below each panel, we also show the uncertainty sizes of the thick blue curves (full coherent sum of resonant contributions), which are computed by propagating the electrocoupling uncertainties via a bootstrap approach, see Ref.~\cite{Blin:2019fre} for details.}
\label{F:F2sing}
\end{figure*}

\begin{figure*}[ht]
\includegraphics[width=0.9\textwidth]{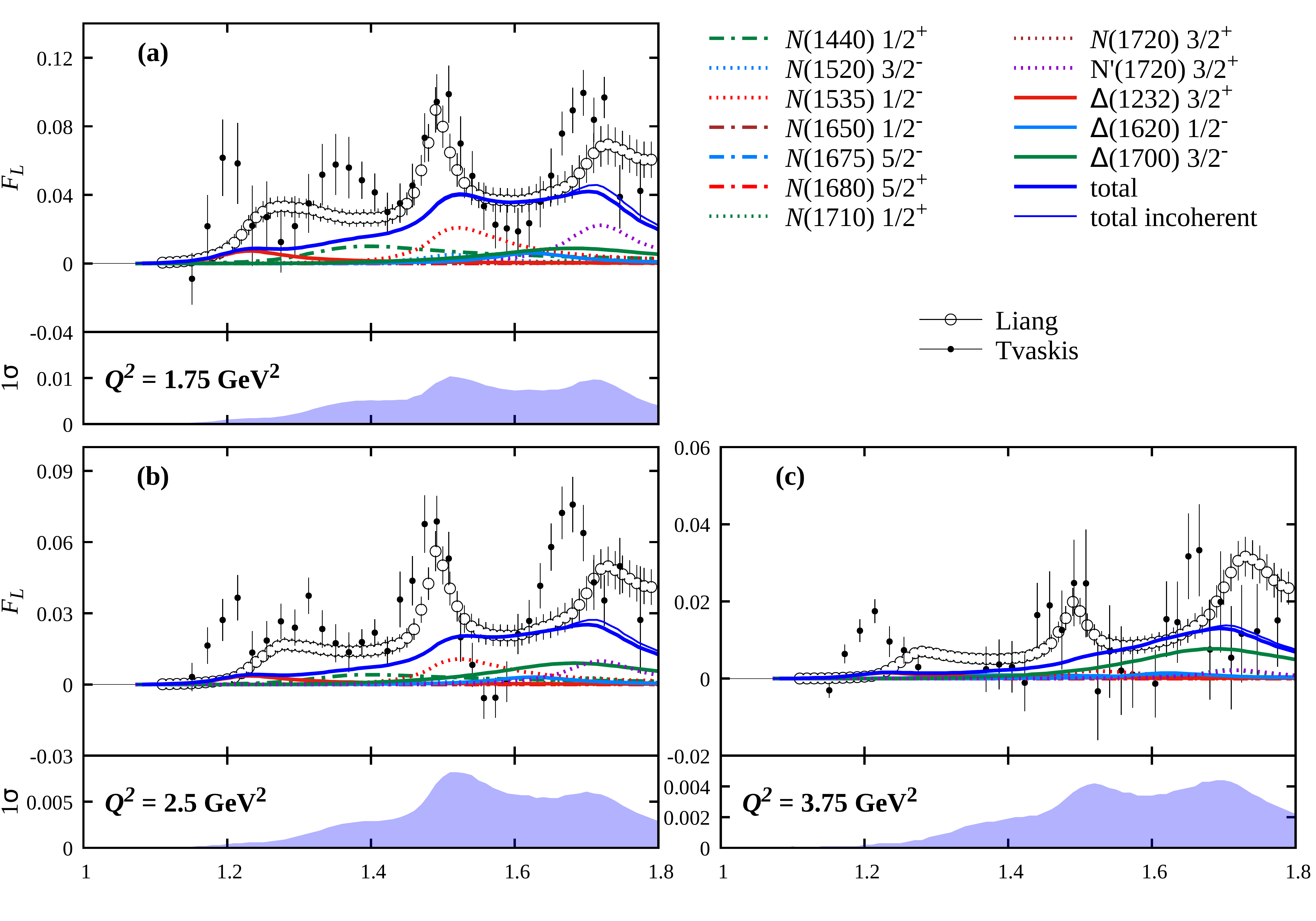}
\caption{Proton $F_L$ structure function at fixed values of $Q^2$ corresponding to the data bins in Ref.~\cite{Tvaskis:2016uxm}:
{\bf (a)}~$Q^2=1.75$~GeV$^2$, 
{\bf (b)}~$Q^2=2.5$~GeV$^2$, 
{\bf (c)}~$Q^2=3.75$~GeV$^2$. The experimental data are from Refs.~\cite{Tvaskis:2016uxm} (filled black circles) and \cite{Liang:2004tj} (open black circles). {The data are compared with the various curves, as  described in Fig.~\ref{F:F2sing}.}}
\label{F:FLsing}
\end{figure*}

Representative examples of the $W$ dependence of $F_2(W,Q^2)$ and $F_L(W,Q^2)$ are shown in Figs.~\ref{F:F2sing} and \ref{F:FLsing} for several values of $Q^2$ between $Q^2 \approx 1$ and 4~GeV$^2$, which lie within the kinematics area over $W$ and $Q^2$ measured in Refs.~\cite{Osipenko:2003bu, Tvaskis:2016uxm}.
Three distinct resonance peaks are clearly seen in the $W$ dependence of both $F_2$ and $F_L$ for the entire $Q^2$ range covered in our analysis, and their qualitative features can be understood from the behavior of the individual resonance contributions displayed in the panels of Figs.~\ref{F:F2sing} and \ref{F:FLsing}.

More specifically, in the first resonance region, the contributions from the $\Delta(1232)\,3/2^+$ resonance to the $F_2$ structure function decrease rapidly with $Q^2$, so that at $Q^2>2$~GeV$^2$ the tail from the $N(1440)\,1/2^+$ state becomes essential.
The relative $\Delta(1232)\,3/2^+$ contribution to the $F_L$ structure function is much smaller than for $F_2$, and falls steeply with $Q^2$.

In the second resonance region, the $N(1520)\,3/2^-$ and $N(1535)\,1/2^-$ states are responsible for the largest contributions to $F_2$, and at $Q^2 \approx 1$~GeV$^2$ they are of comparable size. 
As $Q^2$ increases, the contribution from the $N(1535)\,1/2^-$ becomes dominant.
The $F_L$ structure function in the second resonance region at $Q^2<3$~GeV$^2$ is determined mostly by the contribution from the $N(1535)\,3/2^-$ state.
As $Q^2$ increases above 3~GeV$^2$, the tail from $\Delta(1700)\,3/2^-$ becomes the main resonant contribution in the second resonance region.

The resonance peak in the $F_2$ structure function in the third resonance region is generated by contributions from several nucleon resonances, the biggest impact stemming from the $N'(1720)\,3/2^+$ state.
The $N(1680)\,5/2^+$ and $N(1720)\,3/2^+$ resonances give sub-leading contributions to $F_2$ in this region. 
Because of the intricate interplay with other resonances, the evolution with $Q^2$ of the third resonance peak seen in the $W$ dependencies of $F_2$ becomes rather involved. 
In the range $Q^2 < 3$~GeV$^2$, the contribution from $N(1720)\,3/2^+$ decreases with $Q^2$, the $N(1680)\,5/2^+$ and $N'(1720)\,3/2^+$ resonances become the most significant ones, and the contribution from the $\Delta(1700)3/2^-$ state remains almost negligible in comparison. 
At $Q^2 > 3$~GeV$^2$, the relative contribution from the $\Delta(1700)\,3/2^-$ resonance increases and becomes comparable with the contributions from the $N(1680)\,5/2^+$ resonances at $Q^2 \approx 4$~GeV$^2$.

The behavior of the $F_L$ structure function in Fig.~\ref{F:FLsing} in the third resonance region is determined mostly by the $\Delta(1700)\,3/2^-$ and $N'(1720)\,3/2^+$ resonances.
As was the case for $F_2$, the new $N'(1720)\,3/2^+$ state also plays an important role in the resonant contributions to $F_L$ in this region, and the contribution from $\Delta(1700)\,3/2^-$ dominates the resonant part at $Q^2 \sim 4$~GeV$^2$.
This behavior of the $\Delta(1700)\,3/2^-$ resonance, for both the $F_2$ and $F_L$ structure functions, suggests that further insight can be gained into its structure in the range of high $Q^2 > 4$~GeV$^2$, which will be covered in future nucleon resonance studies with the CLAS12 detector~\cite{Carman:2020qmb, Brodsky:2020vco}.

As should be clear from the above discussion, the $F_2$ and $F_L$ structure functions are sensitive to different combinations of contributions from individual resonances, and studies of both $F_2$ and $F_L$ offer complementary information on the resonant contributions to inclusive electron-proton scattering.
The resonant contribution to $F_2$ in the second resonance region, for example, decreases with $Q^2$ much more slowly than in the first and third resonance regions.
The reason for this is the rather flat $Q^2$ dependence of the transverse electrocoupling of the $N(1535)1/2^-$ in comparison with other resonances, which leads to significant resonant contributions even at large $Q^2$. 
Such behavior is not reflected in the longitudinal $F_L$ structure function (see Fig.~\ref{F:FLsing}), for which the resonance contributions are associated with the longitudinal electrocouplings.
Instead, here one observes that the third resonant peak becomes dominant as $Q^2$ increases.

The effect of the interference between different resonances can be seen in Figs.~\ref{F:F2sing} and \ref{F:FLsing} as the differences between resonant contributions computed within the coherent and incoherent {\it ans\"{a}tze}, as described in Sec.~\ref{sec:res_eval}.
With the exception of the third resonance region, the interference effects are almost negligible, with most vanishing in inclusive observables.
However, the interference effects are clearly seen in the third resonance region in the $W$ dependence of both $F_2$ and $F_L$, mostly due to the interference between the $N'(1720)\,3/2^+$ and $N(1720)\,3/2^+$ resonant amplitudes, but also with a contribution from the interference between the $N(1535)\,1/2^-$ and $N(1650)\,1/2^-$ states. 
Since the transverse electrocouplings $A^R_{1/2}$ and $A^R_{3/2}$ of these pairs of resonances have the same signs, the interference leads to an {\it enhancement} of the resonant contribution to $F_2$.
In contrast, the longitudinal electrocouplings $S^R_{1/2}$ of the $N'(1720)\,3/2^+$ and $N(1720)\,3/2^+$ states have opposite signs, leading to destructive interference and {\it suppression} of the resonant contributions to $F_L$.   

%............................................................................
\subsection{$Q^2$ evolution of resonant contributions}

\begin{figure*}[t]
\includegraphics[width=0.9\textwidth]{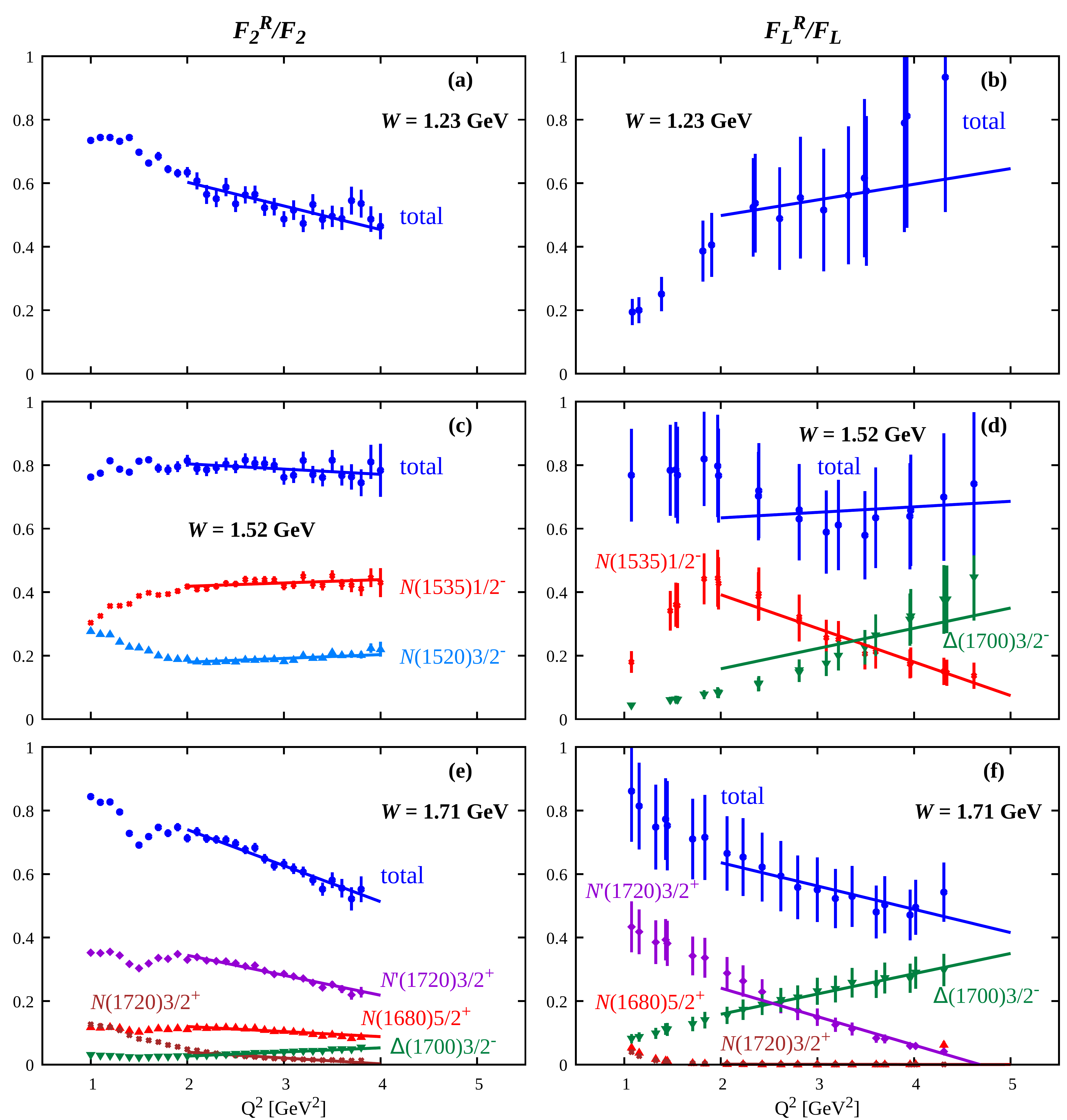}
\caption{$Q^2$ evolution of ratios of the resonant contributions to the total inclusive structure function, $F^R_2/F_2$ (left) and $F^R_L/F_L$ (right), in 
{\bf (a), (b)} the first ($W=1.23$~GeV), 
{\bf (c), (d)} the second ($W=1.52$~GeV), and 
{\bf (e), (f)} third ($W=1.71$~GeV) resonance regions.
In the second and third resonance regions, the ratios are also shown for individual resonances with the largest contributions. The ratios are fitted by linear functions (solid lines) for $Q^2 > 2$~GeV$^2$.}
\label{F:F2Q2evo}
\end{figure*}

In order to study the evolution of the resonance structure with $Q^2$, we evaluate the ratios between the resonant contributions to the $F_2$ and $F_L$ structure functions and the total inclusive structure functions, $F^R_2/F_2$ and $F^R_L/F_L$, at three representative $W$ values corresponding to the first, second and third resonance regions.
In Fig.~\ref{F:F2Q2evo} the $Q^2$ dependence of the ratios is shown between $Q^2=1$ and roughly 5~GeV$^2$ (or up to where data are available), together with a linear fit to the ratios for $Q^2 > 2$~GeV$^2$.
The resonant contributions to $F_2$ and $F_L$ remain significant over the entire $Q^2$ range considered, accounting for 40\%--50\% in the first and third resonance regions, even up to $Q^2 \sim 4$~GeV$^2$, and around 75\% in the second resonance region. 
% {\color{blue}
Analyses of exclusive meson electroproduction data from Jefferson Lab Hall~B~\cite{Aznauryan:2009mx, Mokeev:2012vsa, Mokeev:2015lda, Denizli:2007tq, Park:2014yea} demonstrated that the $\gamma^* p N^*$ can be extracted with an accuracy better than 15\%, when the resonant contributions to the exclusive meson electroproduction cross sections are above 10\%. 
Therefore, the considerable relative size of the resonance contributions to the inclusive structure functions observed in our analysis at $Q^2$ up to 4-5~GeV$^2$, suggests promising prospects for extracting the $\gamma^{*}pN^*$ from the future exclusive meson electroproduction data with the CLAS12 detector at $Q^2>4$~GeV$^2$~\cite{Carman:2020qmb, Brodsky:2020vco}. 

Interestingly, the ratios $F^R_2/F_2$ in the first and third resonance regions are observed to decrease with $Q^2$, and at $Q^2 > 2$~GeV$^2$ their evolution with $Q^2$ is well described by linear functions with similar slope values, of the order of $-0.1$ GeV$^{-2}$.
In contrast, the $F^R_2/F_2$ ratio in the second resonance region remains nearly $Q^2$ independent at $Q^2 > 2$~GeV$^2$, suggesting that both resonant and nonresonant contributions here decrease with $Q^2$ at the same rate.
Furthermore, the leading $N(1535)\,1/2^-$ and $N(1520)\,3/2^-$ contributions are also mostly independent of $Q^2$ for $Q^2 > 2$~GeV$^2$, revealing that contributions from both these resonances decrease with $Q^2$ at the same rate as the nonresonant contributions. 
The underlying cause for these correlations between the $Q^2$ dependence of the resonances in the second resonance region remains an interesting question.

The large contribution from the new baryon state, $N'(1720)\,3/2^+$, at $Q^2 > 2$~GeV$^2$ represents the driving feature in the $Q^2$ evolution of the $F^R_2/F_2$ ratio in the third resonance region. 
Future studies with CLAS12~\cite{Carman:2020qmb} of the resonances in the third resonance region will elucidate the particular structural features of the new $N'(1720)\,3/2^+$ resonance whose contribution underlies a less pronounced decrease with $Q^2$ than that of the regular $N(1720)3/2^+$.

As for the longitudinal structure function ratio $F^R_L/F_L$,
in the first resonance region we fit the ratio only for data with $2 < Q^2 < 3.9$~GeV$^2$, since the bins with larger $Q^2$ have very large uncertainties and we do not consider them reliable. 
There is only a very mild dependence of the ratio in the first resonance region with $Q^2$, with mean value $\approx 0.6$~GeV$^2$.
More accurate data on the $F^R_L/F_L$ ratio are needed to shed light on the $Q^2$ evolution of the competition between the resonant and nonresonant contributions to $F_L$ in the first resonance region.
In the second resonance region, $F^R_L/F_L$ remains $Q^2$ independent for $Q^2 > 2$~GeV$^2$, around $0.6 - 0.7$.
There are two leading contributions to $F_L$ in this region: the $N(1535)\,1/2^-$ resonance, and the tail of the $\Delta(1700)\,3/2^-$ state from the third resonance region.
While the contribution from $N(1535)\,1/2^-$ to $F^R_L/F_L$ decreases with $Q^2$, the contribution from the tail of $\Delta(1700)\,3/2^-$ grows with $Q^2$.
The opposing trends in the $Q^2$ evolution of these two resonances result in a nearly $Q^2$ independent $F^R_L/F_L$ ratio in the second resonance region.

In the third resonance region, the interplay of contributions from four individual resonances gives rise to a ratio $F^R_L/F_L$ that decreases with $Q^2$.
In particular, the contributions from the $N'(1720)\,3/2^+$ and $\Delta(1700)\,3/2^-$ states, which decrease and increase with $Q^2$, respectively, define the $Q^2$ dependence of $F^R_L/F_L$. 
Interestingly, the longitudinal contribution from the $\Delta(1700)\,3/2^-$ resonance reveals a less pronounced fall-off with $Q^2$ than the nonresonant contributions to the inclusive $F_L$.
Whether the same trend for this state persists at $Q^2 > 4$~GeV$^2$ represents an interesting open question for the CLAS12 program.

%%%%%%%%%%%%%%%%%%%%%%%%%%%%%%%%%%%%%%%%%%%%%%%%%%%%%%%%%%%%%%%%%%%%%%%%%%%%%
\section{Quark-hadron duality}
\label{sec:duality}

The availability of the results on electrocouplings from the CLAS electroproduction data, along with the separated inclusive $F_2$ and $F_L$ structure functions, allows us to delve further into the quark hadron duality from comparison between the structure functions in the resonance and DIS regions.
As observed half a century ago in the early SLAC experiments by Bloom and Gilman~\cite{Bloom:1970xb}, and confirmed recently in high precision measurements at Jefferson Lab ~\cite{Niculescu:2000tj, Niculescu:2000tk, Malace:2009kw, Malace:2009dg, Niculescu:2015wka}, the average of the proton structure function measured in the resonance region bears a striking resemblance to the structure function extracted from higher-$W$ (lower-$x$) data and extrapolated to lower $W$. 
In this section we first present a quantitative study of Bloom-Gilman duality for the proton $F_2$ and $F_L$ structure functions using extrapolations from several different global QCD parametrizations, and then examine the $Q^2$ evolution of the truncated moments of the structure functions relative to the experimental data.

%............................................................................
\subsection{Local quark-hadron duality}
\label{pdfinsight}

\begin{figure*}[t]
\includegraphics[width=0.9\textwidth]{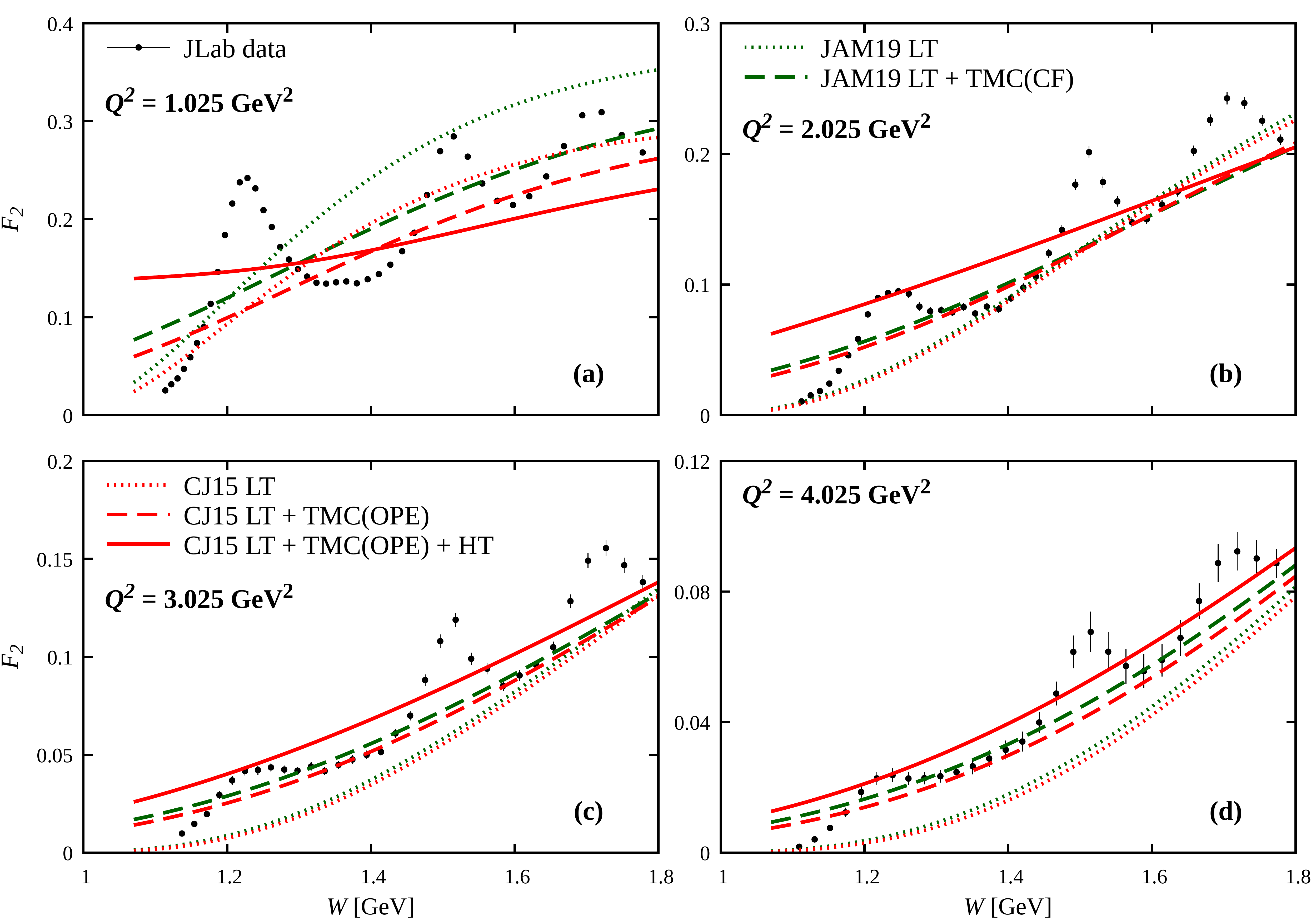}
\caption{Comparison between interpolated $F_2$ structure function data from CLAS~\cite{CLAS:SFDB} (filled black circles) and $F_2$ computed from PDF parametrizations fitted to higher-$W$ data and extrapolated to the resonance region, versus $W$ at fixed values of $Q^2$:
{\bf (a)}~$Q^2=1.025$~GeV$^2$, 
{\bf (b)}~$Q^2=2.025$~GeV$^2$, 
{\bf (c)}~$Q^2=3.025$~GeV$^2$, 
{\bf (d)}~$Q^2=4.025$~GeV$^2$. 
The DIS-based calculations are derived from the CJ15~\cite{Accardi:2016qay} (red lines) and JAM19~\cite{Sato:2019yez} (green lines) PDF parametrizations using leading twist (LT) contributions only (dotted lines), including target mass corrections (TMC) [OPE~\cite{Georgi:1976ve} for CJ15 and CF for JAM19] (dashed lines), and higher twist (HT) contributions [for CJ15 only] (solid line).}
\label{F:F2pdf}
\end{figure*}

% CJ15 = LT + TMC + HT ==> W^2>3 GeV^2
% JAM19 = LT + TMC ==> W^2>10 GeV^2

\begin{figure*}[t]
\includegraphics[width=0.9\textwidth]{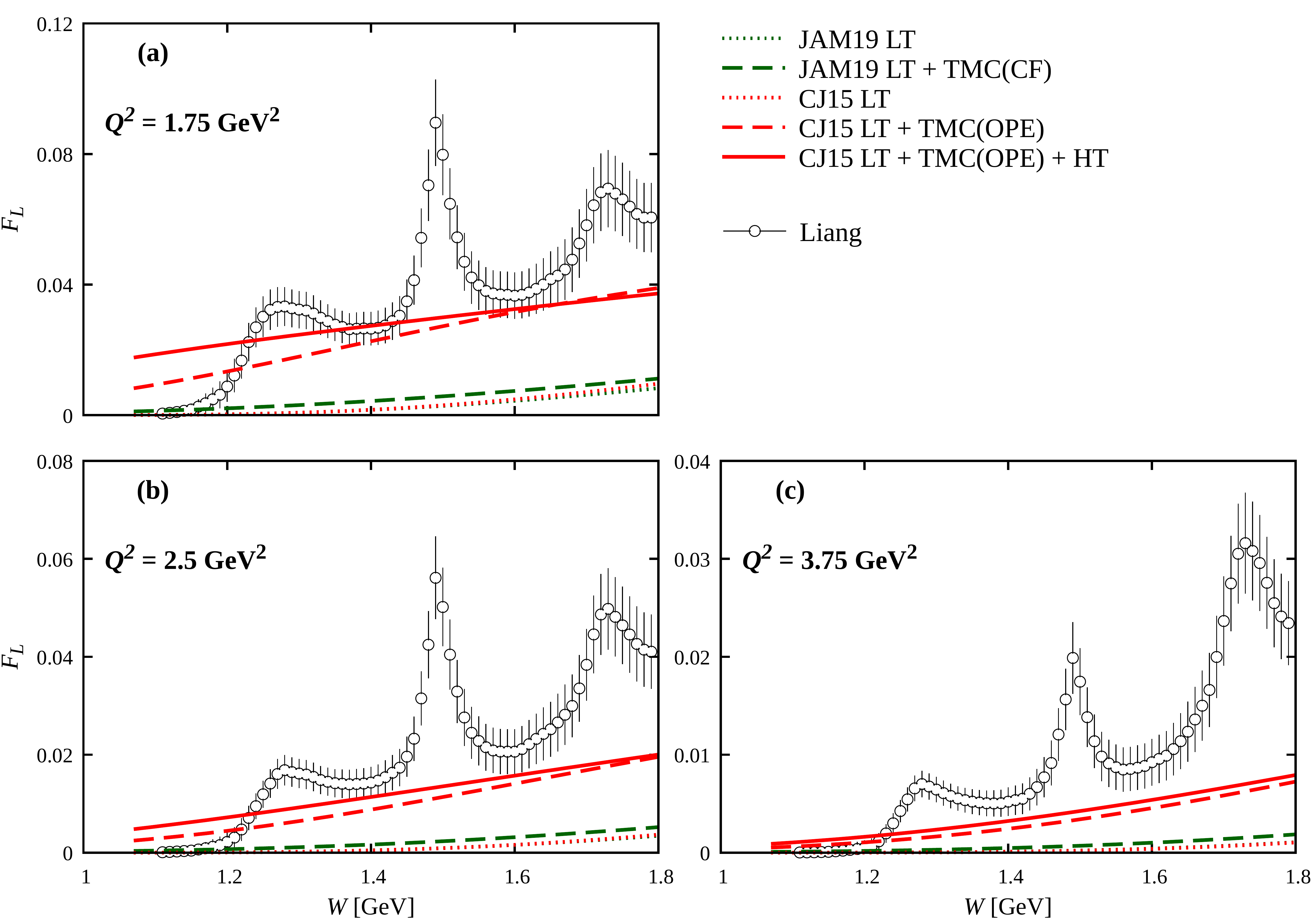}
\caption{As in Fig.~\ref{F:F2pdf}, but for the proton $F_L$ structure function evaluated at 
{\bf (a)}~$Q^2=1.75$~GeV$^2$, 
{\bf (b)}~$Q^2=2.5$~GeV$^2$, 
{\bf (c)}~$Q^2=3.75$~GeV$^2$,
with the data (open circles) taken from Ref.~\cite{Liang:2004tj}.}
\label{F:FLpdf}
\end{figure*}

To explore the details of the duality between the resonance region structure functions and the functions extrapolated from the high-$W$ region, we compare in Fig.~\ref{F:F2pdf} the interpolated inclusive $F_2$ structure function data from CLAS~\cite{CLAS:SFDB} with $F_2$ computed from PDF parametrizations fitted to DIS region data and extrapolated to the resonance region.
The structure functions are shown versus $W$ at fixed values of $Q^2$ between $Q^2 \approx 1$~GeV$^2$ and 4~GeV$^2$.

For the PDF-based calculations, we compute the $F_2$ structure function from next-to-leading order PDF parametrizations from the CJ15~\cite{Accardi:2016qay} and JAM19~\cite{Sato:2019yez} global QCD analyses.
Both of these analyses use similar high-energy scattering data sets, including DIS, Drell-Yan, and weak vector boson production.
The CJ15 analysis~\cite{Accardi:2016qay} includes in addition data at relatively low values of $W$, $W^2 > 3$~GeV$^2$, which requires finite-$Q^2$ corrections, such as target mass corrections (TMCs) and higher twist effects, to be taken into account.
In contrast, the JAM19 analysis~\cite{Sato:2019yez} applies more conservative cuts that exclude lower-$W$ data, taking $W^2 > 10$~GeV$^2$.
Moreover, since the implementation of the TMCs is not unique~\cite{Schienbein:2007gr, Brady:2011uy, Moffat:2019qll}, the CJ15 analysis adopts the OPE based prescription of Georgi and Politzer~\cite{Georgi:1976ve}, while the JAM19 fit employs the collinear factorization (CF) framework throughout, for which the TMCs were evaluated by Aivazis, Olness and Tung (AOT)~\cite{Aivazis:1993kh}.
The CJ15 parametrization also fitted the higher twist, power-suppressed correction to $F_2$ (see Appendix~\ref{app:TMC}), while the JAM19~\cite{Sato:2019yez} analysis was not sensitive to these effects because of the higher-$W$ cuts imposed.

The comparison between the resonance region data, averaged between the resonance peaks and valleys, and the leading twist structure functions, for both the CJ15 and JAM19 PDF parametrizations, shows that overall the extrapolated functions underestimate the data.
After including the TMC effects, the values of $F_2$ generally increase, with the exception of the lowest $Q^2$ value, and show much better agreement with the data, with the differences between $F_2$ with and without the TMCs also decreasing with $W$ across all $Q^2$ bins.
This remains true for both the CJ15 (OPE TMCs) and JAM19 (CF TMCs) results.
For the CJ15 calculation, including the additional higher twist contribution to the structure function, the values of $F_2$ further increase and improve the overall agreement with the data in the resonance region.
An exception is the first resonance region, which is generally overestimated for $Q^2 \gtrsim 2$~GeV$^2$.

This behavior suggests the intriguing possibility of utilizing resonance region data at $Q^2 \gtrsim 2$~GeV$^2$ to provide constraints for nucleon PDFs, something which has been speculated about previously~\cite{Melnitchouk:2005zr} but never implemented in practice.
At $Q^2 \lesssim 2$~GeV$^2$, the different PDF parameterizations and the TMC prescriptions result in a more substantial spread in the predicted behavior of $F_2$ in the resonance region, which cautions that either quark-hadron duality or the perturbative expansion may not be valid here.

For the longitudinal structure function $F_L$ in Fig.~\ref{F:FLpdf}, the comparison between the resonance region data and the PDF-based extrapolations shows a much greater dependence on the finite-$Q^2$ prescriptions.
While the leading twist part of $F_L$ is almost identical for the CJ15 and JAM19 analyses, the OPE implementation of TMCs in the former gives a stronger effect than in the CF implementation, which JAM19 uses.
These differences can be understood from Eqs.~(\ref{eq.TMC-CF}) and (\ref{eq.TMC-OPE}) and, in particular for the OPE formulation, the target mass corrected $F_L$ structure function receives a contribution at order $\xb^2 M^2/Q^2$ which is proportional to an integral over the (large) $F_2$ structure function.
Such a contribution is not present in the CF formulation, (\ref{eq.TMC-CF}), so that the TMC effect in the OPE prescription is considerably larger.
For the CJ15 analysis, the addition of the higher twist component increases $F_L$, but does not affect the general features.

This comparison suggests that $F_L$ is dominated by resonant contributions and the interference between the resonant and nonresonant components in the entire range of $Q^2 \lesssim 4$~GeV$^2$ covered in our analysis.
Improvements in the empirical determination of the separated $F_L$ would be helpful in further clarifying the role of the subleading contributions.

While the similarity of the average resonance region data with the structure functions extrapolated from higher $W$ suggests hints for a deeper connection between nonperturbative resonance and partonic physics, in order to establish whether this can be utilized in practice to learn about PDFs from resonance region data requires this relation to be established at a more quantitative level.
In the next section we discuss how one may better assess the degree to which the duality holds using a quantitative framework based on truncated moments of structure functions.

%...........................................................................
\subsection{Truncated moments of structure functions}
\label{ssec:global}

Since a smooth, perturbative QCD-based function cannot hope to describe the detailed structures of resonance peaks and valleys, any manifestation of duality in electron--proton scattering must involve some averaging over resonances or intervals in $W$ in the resonance region.
Formally, in the language of the OPE, the appearance of duality can be understood in terms of suppression of higher twist contributions to moments of the structure functions, integrating over the entire range of $x$ from 0 to 1~\cite{DeRujula:1976baf, Ji:1994br}.
A practical limitation of this is that the computation of full moments can only ever be approximated by inclusion of extrapolations into unmeasured regions of $x$ or $W$.

\begin{figure*}[t]
\includegraphics[width=0.9\textwidth]{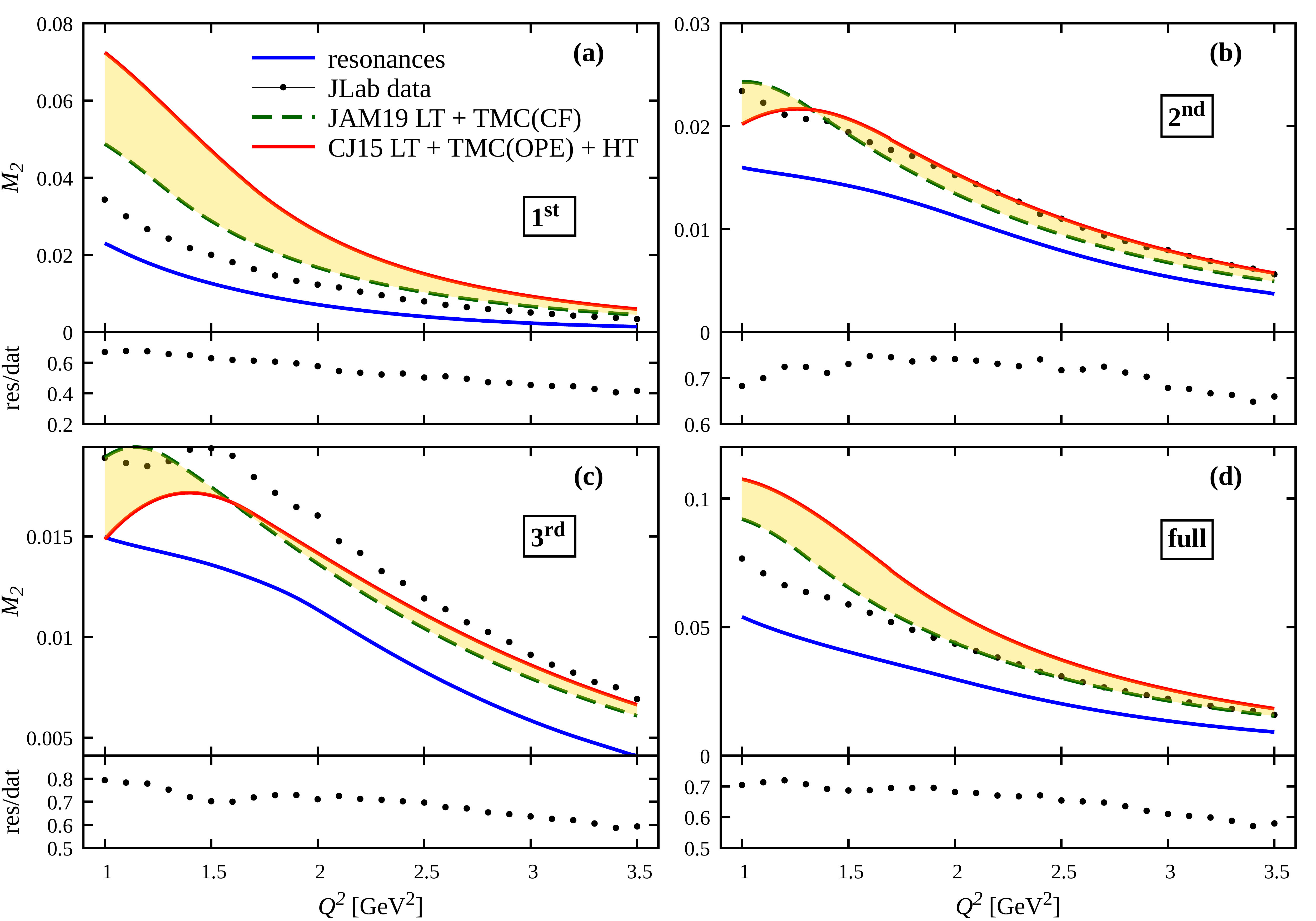}
\caption{Truncated moments $M_2$ of the $F_2$ structure function versus $Q^2$ for the 
{\bf (a)} first, or $\Delta(1232)$, 
{\bf (b)} second, and
{\bf (c)} third
resonance regions, as well as 
{\bf (d)} the full resonance region from the pion threshold to $W=1.75$~GeV. The moments from the experimental results~\cite{CLAS:SFDB} (black circles, with uncertainties smaller than the circle sizes) are compared with the resonant contributions (blue lines) and the structure function moments computed from the JAM19~\cite{Sato:2019yez} (green lines) and CJ15~\cite{Accardi:2016qay} (red lines) PDFs, with the latter including also higher twist terms. The yellow bands between the CJ15 and JAM19 parametrizations reflect the systematic theoretical uncertainties. Also shown beneath each panel is the ratio of the resonant contributions to the data in each $W$ region.}
\label{F:F2trunc}
\end{figure*}

\begin{figure*}[t]
\includegraphics[width=0.9\textwidth]{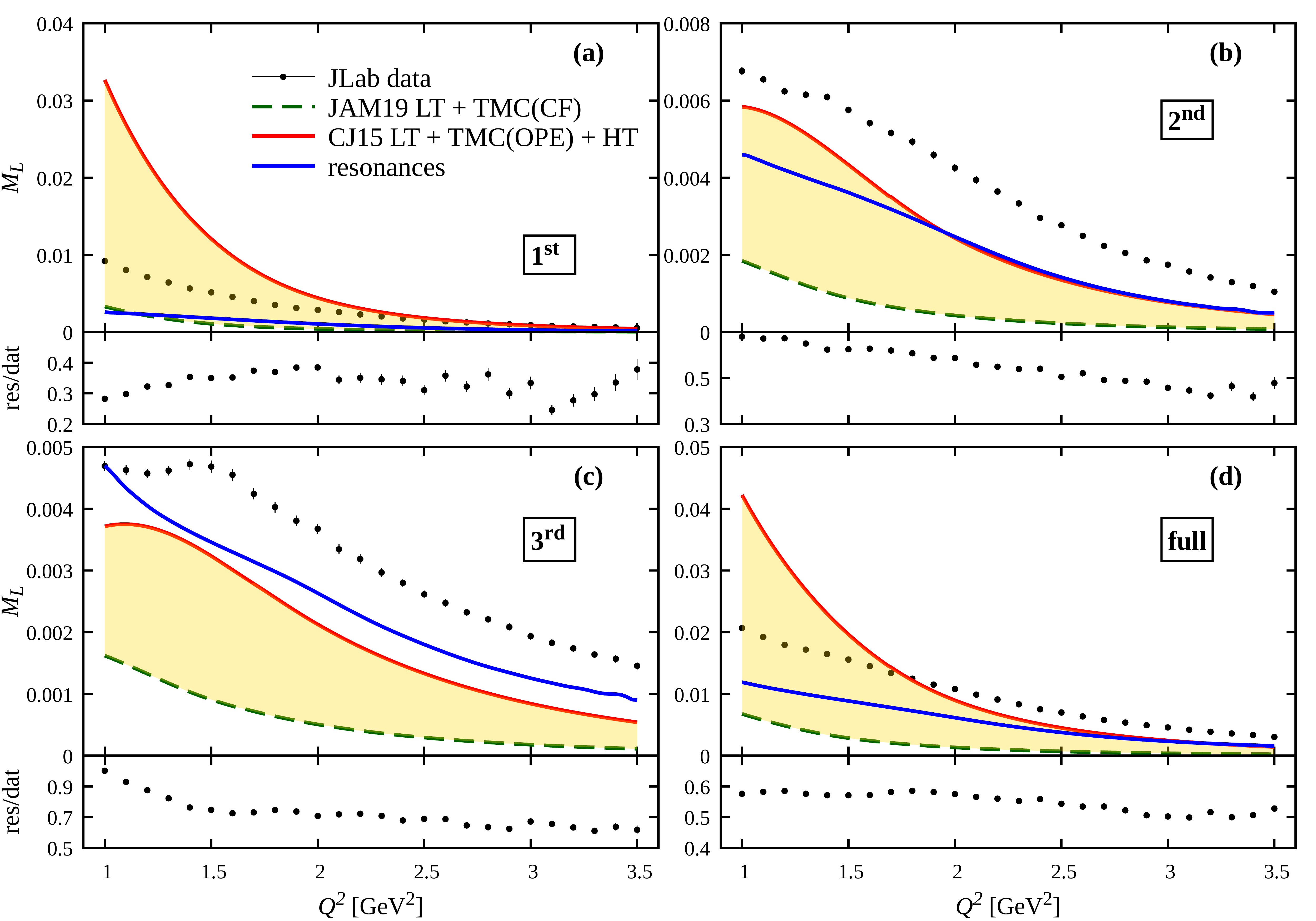}
\caption{As in Fig.~\ref{F:F2trunc}, but for the truncated moments $M_L$ of the $F_L$ structure function. The data (black circles) are extracted from the experimental CLAS results~\cite{CLAS:SFDB} and the Hall~C $\RLT$ ratios~\cite{Liang:2004tj}.}
\label{F:FLtrunc}
\end{figure*}

To avoid introducing uncontrolled assumptions into the analysis, it may be preferable to quantify duality by considering truncated moments, whose evaluation can then be entirely data driven~\cite{Psaker:2008ju}.
We define the lowest truncated moments of the $F_2$ and $F_L$ structure functions in an interval $\Delta \xb \equiv x_{\rm max} - x_{\rm min}$ at a fixed $Q^2$ value by
\begin{align}
\label{moments}
    M_{2,L}(x_{\rm min}, x_{\rm max}; Q^2)
    = \int_{x_{\rm min}}^{x_{\rm max}} \mathrm{d}\xb\, F_{2,L}(x,Q^2).
\end{align}
The truncated moments are evaluated from the pion production threshold, \mbox{$W_\pi = M + m_\pi$}, to the maximal $W$ value of 1.75~GeV where % experimental 
{the results on the $\gamma^* p N^*$ electrocouplings} are currently available~\cite{Carman:2020qmb}, as well as in the $W$ intervals
    $W_\pi < W < 1.38$~GeV,
    $1.38  < W < 1.58$~GeV,
and $1.58  < W < 1.75$~GeV, corresponding to the first, second, and third resonance regions.
The integration limits in Eq.~(\ref{moments}) $[x_\text{min},x_\text{max}]$ correspond to these $W$ intervals at a given value of $Q^2$.

For the truncated moments of the interpolated experimental data on $F_{2,L}$~\cite{CLAS:SFDB}, we evaluate these as discrete sums over bins $i$,
\begin{align}
M_{2,L}^{\rm exp} = \sum_i\, \mathrm{d}x_i\, F_{2,L}^i(x_i,Q^2),
\end{align}
where $i$ runs over all the bins for which $x_{\rm min} \leq x_i \leq x_{\rm max}$, $F_{2,L}^i$ is the value of the structure function in that bin, and $\mathrm{d}x_i$ is the size of the bin. 
To quantify the degree to which duality holds for the individual regions of $W$, we compare the empirical moments with the moments of structure functions computed from the CJ15~\cite{Accardi:2016qay} and JAM19~\cite{Sato:2019yez} PDFs extrapolated from higher $W$.
Note that the CJ15 results include TMCs (using the OPE prescription) and parametrized higher twist contributions as in Sec.~\ref{ssec.HTs}, while the JAM19 results include TMCs (from the CF prescription) only.
The differences between the truncated moments calculated from the CJ15 and JAM19 parametrizations can be interpreted as reflecting the systematic theoretical uncertainties associated with the extrapolations from high $W$ to low $W$.

The results for the $M_2$ and $M_L$ truncated moments are shown in Figs.~\ref{F:F2trunc} and \ref{F:FLtrunc}, respectively, for $Q^2$ between 1~GeV$^2$ and 3.5~GeV$^2$.
For the $M_2$ moments evaluated for the entire resonance region, $W_\pi < W < 1.75$~GeV, there is reasonable agreement, within systematic theoretical uncertainties shown as a yellow band, between the experimental data and the extrapolations from the DIS region for {\color{black} $Q^2 \gtrsim 2$~GeV$^2$}.
A similar agreement is observed in the second resonance region in Fig.~\ref{F:F2trunc}, down to even smaller $Q^2$ values, $Q^2 \gtrsim 1.5$~GeV$^2$.
In the third resonance region the extrapolated results generally underestimate the data by $\sim 10\%-30\%$, while the strongest violation is seen in the first or $\Delta(1232)$ resonance region, where the extrapolated results overestimate the data at all $Q^2$ considered.
The cancellation of the effects in the first and third resonance regions contributes to the observation of duality for the truncated moment of the full resonance region.
Interestingly, the ratio of the resonance contributions to the truncated $M_2$ moments relative to the total decreases with $Q^2$ by less than 20\% across the range of photon virtualities considered in this analysis.

A qualitatively similar behavior was observed for the different resonance regions in the truncated moment analysis by Psaker {\it et al.}~\cite{Psaker:2008ju}, although that work focused only on the $F_2$ structure function moments.
The availability of the more elusive $F_L$ data~\cite{Liang:2004tj} makes possible to test duality also for the truncated moments $M_L$ of the longitudinal structure function.
These are shown in Fig.~\ref{F:FLtrunc} for the same range of kinematics as for the $M_2$ moments.

As observed already in Fig.~\ref{F:FLpdf} above, for the truncated $M_L$ moments there is a much stronger dependence on the prescription adopted for accounting for the subleading $Q^2$ corrections.
In particular, the role of higher twist effects is significantly greater, as the comparison between the CJ15 and JAM19 extrapolated results shows.
For all kinematics the results with leading twist PDFs supplemented with the CF TMCs in the JAM19 case lie significantly below the data.
However, even with the numerically more important OPE TMCs implemented in the CJ15 analysis, together with the higher twist contributions, the results for the truncated moments of the longitudinal function extrapolated from higher $W$ generally underestimate the data across all $W$ regions (with the exception of the $\Delta(1232)$ region at low $Q^2$, where the CJ15 result lies above the data).

The conclusion to be drawn from this comparison is that either duality is strongly violated in the longitudinal channel, or a better understanding of the leading twist and higher twist contributions to $F_L$ is needed. 
Ideally, one would like to perform a simultaneous analysis of the DIS and resonance regions, using a combination of perturbative QCD and phenomenological tools for the most self-consistent analysis.

As for the $M_2$ moments case, the ratio of the resonance contributions to the truncated $M_L$ moments relative to the total varies weakly with $Q^2$ for $Q^2 \gtrsim 2$~GeV$^2$.
This suggests a similar $Q^2$ behavior of the resonant and nonresonant background contributions to the structure function.
The result also makes it promising to study the nucleon resonance $\gamma^* p N^*$ at $Q^2 \gtrsim 4$~GeV$^2$ with CLAS12~\cite{Carman:2020qmb, Brodsky:2020vco}. 

%%%%%%%%%%%%%%%%%%%%%%%%%%%%%%%%%%%%%%%%%%%%%%%%%%%%%%%%%%%%%%%%%%%%%%%%%%%%
\section{Summary and outlook}
\label{sec:outlook}

To summarize the results presented in this paper, we have computed the nucleon resonance contributions to the inclusive proton $F_2$ and $F_L$ structure functions from the resonance electroexcitation amplitudes available from exclusive meson electroproduction data with CLAS~\cite{Carman:2020qmb, Blin:2019fre}.
This has enabled us to systematically study the $Q^2$ dependence of the resonance contributions in the mass range up to 1.75~GeV and quantify the phenomenon of quark-hadron duality in inclusive electron-proton scattering.
The resonant contributions to the structure functions is evaluated using an update of the {\it ansatz} in Ref.~\cite{Blin:2019fre}, which allows interference effects from different nucleon resonances to be taken into account.
The experimental results on the inclusive structure functions have also been updated by employing empirical $\RLT$ results obtained in the resonance region from the measurements in Hall~C at Jefferson Lab~\cite{Liang:2004tj,Tvaskis:2016uxm}.

We found that each of the three resonance peaks observed in the $W$ dependence of $F_2$ is reproduced by the contributions from several excited states of the nucleon. 
The $\Delta(1232)\,3/2^+$ resonance is the dominant contribution to the peak in the first resonance region, while the contributions from the $N(1520)\,3/2^-$ and $N(1535)\,1/2^-$ resonances combined dominate the resonance structure in the second resonance region.
The peak in the third resonance region is predominantly a combination of the $N(1680)\,5/2^+$, $N(1720)\,3/2^+$, and $N'(1720)\,3/2^+$ resonances.
For $Q^2 > 3$~GeV$^2$ the contribution from the  $\Delta(1700)\,3/2^-$ state also becomes sizeable, while the contribution from the $N(1720)\,3/2^+$ diminishes.

As $Q^2$ increases, in each of the three prominent resonance regions the contributions from the tails of the resonances located in other resonance regions, and particularly those with large widths, become more pronounced.
The first resonance region is affected by the tails from the $N(1440)\,1/2^+$ state, while the second resonance region receives  contributions from tails of the $N(1440)\,1/2^+$ and resonances in the third resonance region.
The resonance structure in the third resonance region is affected by the tails mainly from the $N(1440)\,1/2^+$ and $N(1535)\,1/2^-$ states.
This observation emphasizes the need to account for contributions from all prominent resonances in the extraction of the $\gamma^* p N^*$ electrocouplings from exclusive meson electroproduction data.

The data on $F_L$ offer information on the resonant contributions complementary to $F_2$.
Since $F_L$ is sensitive to the resonance electroexcitation by longitudinally polarized virtual photons, the resonant contributions are generated by different combinations of resonance electrocouplings than those in $F_2$. 
In particular, the peak in the second resonance region is generated by the leading contribution from $N(1535)\,1/2^-$, and we observe a transition to the leading contribution from the tail of $\Delta(1700)\,3/2^-$ as $Q^2$ increases.
Data on both $F_2$ and $F_L$ are therefore of particular importance in gaining insight into the resonant contributions to inclusive electron scattering.

The results on the $Q^2$ evolution of the resonant contributions to $F_2$ in the first, second, and third resonance regions demonstrate that the resonant part remains sizeable, from $\approx 40\%$ to 80~\%, even at the highest photon virtualities in our analysis, $Q^2 \approx 4$~GeV$^2$.
They suggest promising prospects for the experimental exploration of nucleon resonance electroexcitation at $Q^2 > 4$~GeV$^2$, which is already underway with CLAS12 at Jefferson Lab~\cite{Carman:2020qmb, Brodsky:2020vco}.
In the second resonance region, at $Q^2 > 2$~GeV$^2$ the resonant contribution to $F_2$ remains approximately $Q^2$ independent, suggesting that both the resonant and nonresonant contributions are characterized by a similar rate of fall-off with $Q^2$.
Gaining insight into the strong interaction dynamics that underlie these correlations in the second resonance region represents an important goal in the exploration of the quark-hadron transition~\cite{Brodsky:2020vco}.
The $Q^2$ evolution of the resonant part in the third resonance region finds the largest contribution from the new $N'(1720)\,3/2^+$ resonance, suggesting interesting possibilities for gaining insight into the structure of new, previously ``missing'' resonances.

We also analyse the transition of the inclusive structure functions from the resonance region at $Q^2 < 2$~GeV$^2$, dominated by nonperturbative, long-distance multi-quark correlations, to the resonance and DIS regions at high $Q^2$ where the structure functions are more efficiently described through single parton processes via leading twist PDFs.
In an effort to shed light onto the workings of quark-hadron duality in a quantitative way, we compared the $W$ dependence of the empirical $F_2$ and $F_L$ structure functions in the resonance region with those computed from nucleon PDFs. For the latter, the parameters are fitted to the high-$W$ DIS region data, and the functional forms are then extrapolated to the resonance region.
We considered different PDF parameterizations and theoretical approaches for taking into account the TMCs, studying also the effects of highr twist corrections to the QCD factorization.

Our analysis shows that at $Q^2 > 2$~GeV$^2$ the $W$ dependence of $F_2$ determined from the DIS region data and extrapolated to the resonant region is broadly consistent with the behavior of $F_2$ measured in the resonance region, which generally oscillates around it.
Quantitative agreement does depend, however, on the specific treatment of TMCs adopted and inclusion of phenomenological higher twist effects.
This is especially relevant for the $F_L$ structure function, where the specific implementation of TMCs has a much greater impact.

To better quantify the relation between the resonance structure functions and those extrapolated from the DIS region, we considered specific truncated moments of the $F_2$ and $F_L$ structure functions.
In particular, 
we evaluated the contributions to the truncated moments from the first, second, and third resonance regions, along with the total resonance region up to $W < 1.75$~GeV, and tracked their $Q^2$ evolution up to $Q^2 \approx 3.5$~GeV$^2$.
Within the systematic uncertainties associated with the extrapolation of the leading twist functions into the resonance region, we found general agreement between the truncated moments computed from the data and from the extrapolated functions for the full $W < 1.75$~GeV region at $Q^2 \gtrsim 2$~GeV$^2$. 
More pronounced manifestation of local duality was observed in the second and third resonance regions at $Q^2 \gtrsim 1$~GeV$^2$ and $Q^2 \gtrsim 3$~GeV$^2$, respectively, suggesting that the resonant contributions to $M_2$ decrease with $Q^2$ at nearly the same rate as the moments from the full $F_2$.

Definitive conclusions about the longitudinal truncated moments are more difficult to draw on account of the greater systematic uncertainties associated with the experimental data extraction, and the TMC prescriptions and higher twist contributions to $F_L$.
However, the resonance region data were generally found to be underestimated by the extrapolated structure function moments, especially at larger $Q^2$ values, indicating either stronger violation of duality in the longitudinal channel, or incomplete understanding of the leading twist and higher twist contributions to $F_L$.

One avenue to pursue in future will be to simultaneously describe the inclusive data in the resonance and DIS regions, merging the global QCD analysis of PDFs with the phenomenological fits to the resonance structures, and minimizing the systematic uncertainties associated with identifying resonance versus background contributions.
This may provide further insight into the relationship between the physics of PDFs and nucleon resonances, and how the latter could be used to provide stronger constraints of PDFs at large values of $x$.

On the experimental side, our results motivate extensions of the inclusive electron scattering studies in the resonance region towards $Q^2 > 4$~GeV$^2$, as well as the extraction of the $\gamma^* p N^*$ electrocouplings at high photon virtualities from the exclusive meson electroproduction data~\cite{Brodsky:2020vco}.
Beyond this, a further extension of this work would explore the spin dependence of the exclusive--inclusive duality, applying the methodology developed here to the case of the spin-dependent $g_1$ and $g_2$ structure functions of the nucleon.
In this case no additional information on the electrocouplings would be needed above that required for the spin-averaged structure functions, and the role of interference effects should be more clearly revealed.

%%%%%%%%%%%%%%%%%%%%%%%%%%%%%%%%%%%%%%%%%%%%%%%%%%%%%%%%%%%%%%%%%%%%%%%%%%%%
\begin{acknowledgements}
We thank Nobuo~Sato for providing the NLO structure function calculation code used in our calculations, and D.~S.~Carman, R.~W.~Gothe, K.~Joo, N. Markov, and V.~Mathieu for helpful discussions and communications.
This work was supported by the US Department of Energy contract DE-AC05-06OR23177, under which Jefferson Science Associates, LLC operates Jefferson Lab. AP has received funding from the European Union's Horizon 2020 research and innovation programme under the Marie Sk{\l}odowska-Curie grant agreement No.~754496.
\end{acknowledgements}

\appendix
%%%%%%%%%%%%%%%%%%%%%%%%%%%%%%%%%%%%%%%%%%%%%%%%%%%%%%%%%%%%%%%%%%%%%%%%%%%%%
\section{Finite-$Q^2$ corrections}
\label{app:TMC}

In this appendix we summarize the main formulas used in this analysis for target mass corrections, based on the OPE and CF approaches, and the higher twist parametrization of the $F_2^p$ structure function from the CJ15 global QCD analysis~\cite{Accardi:2016qay}.

%............................................................................
\subsection{Target mass corrections}

Within the standard collinear QCD factorization~\cite{Collins:1989gx}, and in the high energy limit where the nucleon mass $M^2 \ll Q^2$, the inclusive structure functions can be computed from convolutions of nucleon PDFs $f_i(\xi)$ (with parton momentum fraction $0 < \xi <1$) and hard scattering functions, summed over all parton flavors $i$,
\begin{subequations}
\label{strfun_pdf}
\begin{align}
F_1(\xb,Q^2)
&= \sum_i \int_{\xb}^1\frac{\mathrm{d}\xi}{\xi}\, \widehat{\mathcal{F}}_1^i(\xb/\xi,Q^2)\, f_i(\xi),      \\
F_2(\xb,Q^2)
&= \sum_i \int_{\xb}^1\mathrm{d}\xi\,
\widehat{\mathcal{F}}_2^i(\xb/\xi,Q^2)\, f_i(\xi),        \\
F_L(\xb,Q^2)
&= \sum_i \int_{\xb}^1\mathrm{d}\xi\,
\widehat{\mathcal{F}}_L^i(\xb/\xi,Q^2)\, f_i(\xi),
\end{align}
\end{subequations}
where the kernels $\widehat{\mathcal{F}}_j^i$ ($j=1,2,L$) are computed perturbatively to a given order in the strong coupling constant $\alpha_{s}$~\cite{Collins:2011zzd}. 
Corrections to Eqs.~(\ref{strfun_pdf}) appear in the form of power suppressed corrections, ${\cal O}(\Lambda^2/Q^2)$.
{
Note that both the PDFs $f_i$ and the kernels $\widehat{\mathcal{F}}_j^i$ depend on the renormalization scale, which for ease of notation has been suppressed.}

At finite values of $Q^2$, corrections of the order ${\cal O}(M^2/Q^2)$ appear, modifying the massless results in Eqs.~(\ref{strfun_pdf}) with additional kinematical factors.
These were derived by Aivazis, Olness and Tung (AOT)~\cite{Aivazis:1993kh}, and further elaborated more recently by Moffat {\it et al.}~\cite{Moffat:2019qll}.
In analogy with Eqs.~(\ref{strfun_pdf}), the target mass corrected structure functions in the CF approach become
\begin{subequations}
\begin{align}
F_1^{\rm CF}(\xb,Q^2)
&= \sum_i \int_{x_N}^1\frac{\mathrm{d}\xi}{\xi}\, 
\widehat{\mathcal{F}}_1^i(x_N/\xi,Q^2)\, f_i(\xi),             \\
F_2^{\rm CF}(\xb,Q^2)
&= \frac{(1+\rho)}{2\rho^2} \sum_i \int_{x_N}^1\mathrm{d}\xi\, \widehat{\mathcal{F}}_2^i(x_N/\xi,Q^2)\, f_i(\xi),             \\
F_L^{\rm CF}(\xb,Q^2)
&= \frac{(1+\rho)}{2} \sum_i \int_{x_N}^1\mathrm{d}\xi\, \widehat{\mathcal{F}}_L^i(x_N/\xi,Q^2)\, f_i(\xi),
\end{align}
\label{eq.TMC-CF}
\end{subequations}
where $x_N$ is the Nachtmann scaling variable,
\begin{align}
x_N &= \frac{2\xb}{1+\rho},
\end{align}
with
\begin{align}
\rho^2 &= 1 + \frac{4\xb^2 M^2}{Q^2}.
\end{align}
In the massless limit, $x^2 M^2/Q^2 \to 0$, so that $\rho \to 1$, and the Nachtmann variable approaches the Bjorken scaling variable, $x_N \to x$.
Historically, even before the formulation of the QCD factorization theorems, TMCs were computed in the framework of the OPE by Georgi and Politzer~\cite{Georgi:1976ve}.
Here the target mass corrected structure functions can be written in terms of the massless limit structure functions evaluated at $x_N$ rather than at $x$,
\begin{subequations}
\label{E:FOPE}
\begin{align}
F_1^\text{OPE}(\xb,Q^2)
&= \frac{(1+\rho)}{2\rho}\, F_1(x_N,Q^2)                              \notag\\
&\hspace*{-1.7cm}
+  \frac{\rho^2-1}{4\rho^2}
   \left[ h_2(x_N,Q^2)+\frac{\rho^2-1}{2x\rho}g_2(x_N,Q^2) \right], \\
F_2^\text{OPE}(\xb,Q^2)
&= \frac{(1+\rho)^2}{4\rho^3}\, F_2(x_N,Q^2)                        \notag\\
&\hspace*{-1.7cm}
+  \frac{3x(\rho^2-1)}{2\rho^4}
   \left[h_2(x_N,Q^2)+\frac{\rho^2-1}{2x\rho}g_2(x_N,Q^2)\right],   \\
F_L^\text{OPE}(\xb,Q^2)
&= \frac{(1+\rho)^2}{4\rho}\, F_L(x_N,Q^2)                        \notag\\
&\hspace*{-1.7cm}
+  \frac{x(\rho^2-1)}{\rho^2}
   \left[h_2(x_N,Q^2)+\frac{\rho^2-1}{2x\rho}g_2(x_N,Q^2)\right],
\label{eq.FLope}                                                  \\
\intertext{where the higher order correction terms are given by the integrals}
h_2(x_N,Q^2) &= \int_{x_N}^1\mathrm{d}u\, \frac{F_2(u,Q^2)}{u^2},   \\
g_2(x_N,Q^2) &= \int_{x_N}^1\mathrm{d}u\, (u-x_N)\frac{F_2(u,Q^2)}{u^2}.
\end{align}
\label{eq.TMC-OPE}%
\end{subequations}
Note that for the $F_L^{\rm OPE}$ structure function, the ${\cal O}(M^2/Q^2)$ correction term $h_2$ in Eq.~(\ref{eq.FLope}) involves the $F_2$ structure function.
Since one generally has $F_2 \gg F_L$ across most kinematics, the TMCs for the $F_L^{\rm OPE}$ structure function can be quite sizeable compared with the CF TMCs, which, as Eqs.~(\ref{eq.TMC-CF}) indicate, do not contain such large terms.
This term is the main reason for the large difference between the effects of TMCs on the $F_L$ structure function in Fig.~\ref{F:FLpdf}, and the corresponding moments in Fig.~\ref{F:FLtrunc}, for the CJ15~\cite{Accardi:2016qay} and JAM19~\cite{Sato:2019yez} parametrizations.

For a more detailed discussion of TMCs, see Schienbein {\it et al.}~\cite{Schienbein:2007gr}.
Other TMC approaches are summarized also by Brady {\it et al.}~\cite{Brady:2011uy}.
A discussion of the differences between the CF and OPE approaches to TMCs can be found in Ref.~\cite{Moffat:2019qll}.

%............................................................................
\subsection{Higher twist corrections}
\label{ssec.HTs}

In addition to the TMCs, other finite-$Q^2$ corrections that need to be taken into account include those associated with matrix elements of higher twist, multi-parton operators, which are typically ${\cal O}(\Lambda^2/Q^2)$ suppressed relative to the leading twist contributions.
These are very difficult to compute from first principles, or even from nonperturbative models, so are often parametrized phenomenologically in data analyses.

In the literature the power corrections to DIS structure functions have been fitted by using either additive~\cite{Alekhin:2017kpj} or multiplicative~\cite{Accardi:2016qay} parametrizations.
In the present analysis, we use the higher twist parametrization from the CJ15 analysis, in which the total $F_2$ structure function is given by the multiplicative form~\cite{Accardi:2016qay}
\begin{align}
F_2^\text{HT}(x,Q^2)
&= F_2^\text{OPE}(x,Q^2)\left(1+\frac{C_\text{HT}(x)}{Q^2}\right),
\end{align}
with the coefficient of the $1/Q^2$ term parametrized by a 3-parameter $x$-dependent function,
\begin{align}
C_\text{HT}(x)
&= h_0 x^{h_1} (1 + h_2 x).
\end{align}
The numerical values for the parameters obtained in the CJ15 global analysis are fitted to be~\cite{Accardi:2016qay}
\begin{align}
h_0 &= -3.2874 \pm 0.26061, \notag\\ 
h_1 &= +1.9274 \pm 0.10524, \\
h_2 &= -2.0701 \pm 0.019888. \notag
\end{align}

%%%%%%%%%%%%%%%%%%%%%%%%%%%%%%%%%%%%%%%%%%%%%%%%%%%%%%%%%%%%%%%%%%%%%%%%%%%%%
\bibliographystyle{apsrev4-1-jpac}
\bibliography{cite}{}

\end{document}